\documentclass{jfm}
\usepackage{graphicx}
\usepackage{epstopdf, epsfig}

\usepackage{amsmath}
\usepackage{amssymb}
\usepackage{xcolor}
\usepackage{mathtools}

\newcommand{\uvmax}{|\overline{u'v'}|_{\max}}

\usepackage{lineno}
\usepackage{setspace}
\usepackage{tikz}

\newcount\ndots
\def\drwln#1#2{\raise 2.5pt\vbox{\hrule width #1pt height #2pt}}
\def\spc#1{\hskip #1pt}
\def\solid{\drwln{24}{1.0}\ }

\def\dashed{\hbox {\drwln{4}{1.0}\spc{2}
    \drwln{4}{1.0}\spc{2}\drwln{4}{1.0}}\nobreak\ }

\def\dotted{\hbox {\drwln{2}{1.0}\spc{2}
    \drwln{2}{1.0}\spc{2}\drwln{2}{1.0}}\nobreak\ }

\definecolor{ao}{rgb}{0.0, 0.5, 0.0}

\shorttitle{Wall-pressure spectral model for non-equilibrium boundary layers}

\shortauthor{S. Pargal et al.}

\author{Saurabh Pargal\aff{1,2},
  Junlin Yuan\aff{1} \corresp{\email{junlin@msu.edu}}
 \and Stephane Moreau\aff{2}}

\affiliation{\aff{1}Michigan State University, Michigan, USA
\aff{2}Universit{\'e} de Sherbrooke, Quebec, Canada}

\title{A generalized wall-pressure spectral model for non-equilibrium boundary layers}

\begin{document}
\maketitle

%\linenumbers

\begin{abstract}
This study uses high-fidelity simulations (DNS or LES) and experimental datasets to analyse the effect of non-equilibrium streamwise mean pressure gradients (adverse or favourable), including attached and separated flows, on the statistics of boundary layer wall-pressure fluctuations. The datasets collected span a wide range of Reynolds numbers ($Re_\theta$ from 300 to 23,400) and pressure gradients (Clauser parameter  from $-0.5$ to 200). The datasets are used to identify an  optimal set of variables to scale the wall pressure spectrum: edge velocity, boundary layer thickness, and the peak magnitude of Reynolds shear stress. 
Using the present datasets, existing semi-empirical models of wall-pressure spectrum are shown unable to capture  effects of strong, non-equilibrium adverse pressure gradients, due to inappropriate scaling of wall pressure using wall shear stress, calibration with limited types of flows, and  dependency on model parameters based on friction velocity, which reduces to zero at the detachment point. To address these short-comings, a generalized wall-pressure spectral model is developed with parameters that characterize  the extent of the logarithmic layer and the strength of the wake. 
Derived from the local mean velocity profile, these two parameters inherently carry effect of the Reynolds number, as well as those of the non-equilibrium pressure gradient and its history. Comparison with existing models shows that the proposed model behaves well and is more accurate  in strong-pressure-gradient flows and in separated-flow regions. 

\end{abstract}

\begin{keywords}
%Authors should not enter keywords on the manuscript, as these must be chosen by the author during the online submission process and will then be added during the typesetting process (see \href{https://www.cambridge.org/core/journals/journal-of-fluid-mechanics/information/list-of-keywords}{Keyword PDF} for the full list).  Other classifications will be added at the same time.
\end{keywords}

\section{Introduction}
\label{sec:intro}

The fluctuation in time of the wall pressure beneath a turbulent boundary layer is one of the major sources of flow-induced noise  and  vibrations. Accurate modeling of the statistics of wall-pressure fluctuations is important for noise prediction in a wide range of applications such as wind turbines \citep{avallone2018noise, deshmukh2019wind,venkatraman2023numerical}, cooling fans \citep{Sanjose2018,swanepoel2023experimental, luo2020tip}, propellers \citep{LallierDaniels2021,casalino2021towards}, unmanned/manned air vehicles or drones \citep{lauzon2023aeroacoustics, pargal2023large, celik2021aeroacoustic}, and cabin noise \citep{samarasinghe2016recent, borelli2021onboard}, etc., as well as for prediction of flow-induced structure fatigue \citep{franco2020structural}.
In these applications, the boundary layer flows are often turbulent and non-equilibrium, due to surface curvature and significant pressure gradients that vary in the streamwise direction, which may induce boundary layer separation and can be found in a large range of Reynolds number.
Therefore, the generation of noise in non-equilibrium turbulent boundary layers is physically complex and challenging to model.

The modeling of wall-pressure loading as a noise source  predominantly depends on the power spectral density (PSD) of wall-pressure fluctuations, as well as its spanwise correlation length  and the convection velocity of turbulent structures \citep{amiet1976noise,roger2005back,moreau2009back,lee2021turbulent}.  The focus here is on modeling wall-pressure spectrum (WPS). 
It is established that the WPS of a boundary layer with zero or minimal pressure gradient consists of three ranges \citep{goody2004empirical, farabee1991spectral, chang1999relationship}: (i) a range with $\omega^2$ variation at low frequencies (where $\omega$ is the frequency), (ii) a range with $\omega^{-5}$ behavior at high frequencies, and (iii) an overlap range with a $\omega^{-1}$ decay rate  between the above two ranges. Based on data primarily in equilibrium flows, the width of the overlap range was found to increase with Reynolds number \citep{farabee1991spectral,goody2004empirical}.

Contributions from different layers of  wall turbulence to the WPS has been studied and are summarized below. 
The experimental studies of \cite{farabee1991spectral} suggested different dominant sources for different  wavenumber ranges of the WPS:   the high-wavenumber range is mainly attributed to turbulent activities in the logarithmic region, while the low-wavenumber range is attributed to large-scale turbulent motions in the  outer layer. 
\cite{van2017experimental} quantified the correlations between the fluctuations of wall pressure and those of velocity in different layers of boundary layer    and observed that high-frequency and overlap ranges of the WPS are associated with flows in the buffer and logarithmic regions, respectively.  
%\citet{chang1999relationship} compared the contributions from different velocity source terms (of the Poisson's equation of pressure) to the pressure fluctuations at the wall,  based on a DNS database of low Reynolds number channel flows. They showed that the mean shear (MS) and turbulence-turbulence (TT) terms  are of similar importance. 
%\cite{hu2017simulation} simulated high Reynolds number TBL made similar observations. 
As opposed to earlier studies performed on channel flow or canonical flat-plate boundary layer data, 
\cite{jaiswal2020use} analyzed data collected near the trailing edge of a cambered airfoil with a strong mean adverse pressure gradient (APG) in a highly non-equilibrium turbulent boundary layer, to compare contributions from various velocity sources at different wall-normal locations to the wall-pressure fluctuations based on the pressure Poisson's equation.  They found that the mean shear (MS) term in the inner and logarithmic regions is the  dominant contributor,  especially in the mid-to-high frequency range.

Past studies were mostly on zero-pressure-gradient (ZPG) turbulent boundary layers. They showed that the wall-pressure fluctuations are intensified under a higher Reynolds number, mainly due to the increase in overlap-range spectral contribution.
For a ZPG flat-plate boundary layer, \cite{farabee1991spectral} integrated the pressure spectrum over various frequency ranges  and showed that the low-to-mid frequency range
% ($\omega \delta/u_{\tau}$ $\approx$ 0 to 100 ) 
and the high-frequency range
%  ($0.3Re_{\tau}<\omega\delta/u_{\tau}$) 
were not sensitive to a change in Reynolds number, whereas the significance of the overlap range
%    ($\omega \delta/u_{\tau}$ $\approx 100$  to $0.3Re_{\tau}$ ) 
increases with Reynolds number, leading to an augmentation of $p_{rms}$. 
%Based on these findings, 
They proposed that $p_{rms}^2/\tau_w^2=6.5+1.86 \ln(Re_{\tau}/333)$,
%\begin{eqnarray}
%    \label{eq:Farabee}
%   \frac{p_{rms}^2}{\tau_w^2}={6.5+1.86 \ln(Re_{\tau}/333)}, 
%\end{eqnarray}
which was tested with  ZPG boundary layer and channel flow data. \cite{panton1974wall}  also demonstrated that the overlap range is correlated with the Reynolds number. 

% ==== Effect of pressure gradients:
The current understanding of the WPS in non-zero pressure gradient flows is summarized as follows. Review of the earlier work before the mid 1990s are provided by  \citet{willmarth1975pressure} and \citet{bull1996wall}.
\cite{schloemer1967effects} showed that under an APG low-frequency contents of the WPS become more prominent as large eddies are energized, while the  high-frequency contents become less important. Under a favourable pressure gradient (FPG), however, the opposite applies, with stronger high-frequency contents. The WPS slope in the overlap range also varies with the pressure gradient.
\cite{cohen2018influence} investigated the effects of mild pressure gradient  using large-eddy simulations (LES) and showed scale-based dependencies of the WPS on FPG similar to those observed before.
\cite{na1998structure} conducted direct numerical simulation (DNS) of a boundary layer with prescribed freestream suction and blowing to induce flow separation and reattachment. They  showed that none of the outer, inner or mixed scaling  collapsed the  wall pressure spectra in all regions of the flow.  
Normalization with the local maximum magnitude of the Reynolds shear stress, however, was shown to  collapse the low-frequency range of WPS for APG flows including those with separation \citep{abe2017reynolds,ji2012surface,Caiazzo:JFM:2023}. 

Modeling of turbulent WPS is broadly classified in two categories: (i) semi-empirical modelling and  (ii) analytical modeling based on  solution of the Poisson's equation of pressure \citep{kraichnan1956pressure,panton1974wall, jaiswal2020use, grasso2022advances,palani2023modified, hales2023adapting}. The focus of this paper is on the first approach, which requires a smaller amount of inputs from the flow field in comparison to the analytical modelling approach. 
Existing semi-empirical WPS closures mostly model the magnitude and shape of the WPS normalized by some boundary-layer parameters that are either internal, external or mixed, such as the boundary layer thickness ($\delta$), the edge velocity ($U_e$) and the wall shear stress ($\tau_w=\rho u_\tau^2$, where  $u_{\tau}$ is the friction velocity), etc. For some of these studies see \cite{kraichnan1956pressure}, \cite{corcos1964structure}, \cite{willmarth1975pressure}, \cite{amiet1976noise},\cite{bull1976high},\cite{chase1980modeling}, \cite{goody2004empirical}, \cite{rozenberg2012wall}, \cite{kamruzzaman2015semi}, \cite{lee2018empirical}, \cite{hu2018empirical}, and \cite{pargal2022adverse}.
One of the most widely used semi-empirical models for ZPG boundary layers was proposed by \cite{goody2004empirical}, which captures accurately the Reynolds number effect on the spectrum for these flows. But for boundary layers under pressure gradients, Goody's model may yield large errors, as shown by a number of studies \citep{rozenberg2012wall,kamruzzaman2015semi,catlett2016empirical,hu2013contributions,lee2018empirical, rossi2023prediction}. For instance, in APG flows the more prominent low-frequency contents and reduced mid-to-high-frequency ones of the WPS were not captured by this model.
%\citep{catlett2016empirical,rozenberg2012wall,balantrapu2020wall,fritsch2022fluctuating} in the past. 
\cite{rozenberg2012wall} integrated additional boundary layer flow parameters  to sensitize the model to pressure gradient effects, especially those of APG. The additional parameters include  Clauser's  parameter ($\beta$) \citep{clauser1954turbulent} and Cole's wake parameter ($\Pi$) \citep{coles1956law}. The former includes the local effect of mean  pressure gradients, while the latter represents the cumulative effect of the history of mean pressure gradient up to the considered location in the boundary layer.
Several later models developed modifications of the model that capture effects of other complexities such as wall curvature and FPG. \cite{kamruzzaman2015semi} developed a model by fitting it on a large amount of  experimental WPS data collected in various non-equilibrium  boundary layer flows on airfoils. \cite{hu2018empirical} used the  shape factor ($H$)  and Reynolds numbers ($Re_{\theta}$ or $Re_\tau$) instead of $\beta$ to incorporate the effect of non-equilibrium pressure gradients, as $\beta$ --- a descriptor of local pressure gradient --- does not carry the history effect of a spatially varying pressure gradient.   \cite{lee2018empirical} improved Rozenberg's model based on experimental data gathered from a wide range of flows with different  Reynolds numbers and pressure gradients. \cite{thomson2022semi} proposed a new model for flows with FPG. Recently,   machine learning approaches such as gene expression programming and artificial neural networks were used to model WPS as a function of boundary layer parameters \citep{fritsch2022modeling,dominique2022artificial, shubham2023data, ghiglino2023towards}.

Despite the success of the models mentioned above in the specific flows for which they were developed, these models are not universally applicable to both ZPG flows and those with  non-equilibrium pressure gradients and/or surface curvature, due to the following reasons. (i) Models developed by curve-fitting to data of a limited type of flows do not naturally apply to other flows, such as Goody's model, which  works for ZPG flows only. (ii) Normalizations of wall pressure statistics used for ZPG flows (e.g. $\tau_w$) may not be appropriate for strong-APG flows (e.g. a boundary layer close to separation where $\tau_w$ approaches zero). (iii) Local boundary layer parameters do not  account sufficiently for the history effect of the pressure gradient. In addition, some existing models were developed based on experimental wall-pressure measurements that are supplemented with low-fidelity flow-field data, such as those estimated from XFOIL~\citep{DrelaG89}.
% estimated from XFOIL or Reynolds-averaged Navier-Stokes (RANS) predictions. 

The objective of this %paper 
study is therefore to develop a general WPS model that is tunable for both ZPG and non-equilibrium, strong-pressure-gradient turbulent boundary layers, as well as special cases such as flow separation and reattachment. To this end,  model parameters that derive from the local mean velocity profile are  incorporated to sensitize the model to the streamwise pressure gradient and its history. An appropriate pressure normalization for flows with and without pressure gradients is used. The model is calibrate based on a large and inclusive database,  containing both  experimental measurements and  DNS/LES data (existing or new)  of flows over a wide range of Reynolds number, with or without separation. 

The organization of the paper is as follows.
Section~\ref{sec:datasets} describes the database,  
Section~\ref{sec:Blayer} presents the boundary layer development of the cases in the datasets,  
Section~\ref{sec:p_statistics} discusses the wall-pressure fluctuations and WPS in the datasets,  
Section~\ref{sec:wps} discusses the performances of existing WPS models and introduces a new generalized WPS model, and conclusions are presented in Section~\ref{sec:conclusions}.

\section{Datasets collection}
\label{sec:datasets}
The first step to develop a generalized WPS model  is to collect and analyze high-fidelity datasets in a wide range of flows. 
%In this section, the database collection from  simulations (DNS and LES) and experiments is  discussed. 
The goal is to collect datasets for both equilibrium and non-equilibrium boundary layers, including ZPG, FPG, and APG flows, with or without wall  curvature (as in boundary layers developed on airfoils) and  boundary layer separation and reattachment, across a wide range of Reynolds number based on momentum thickness ($Re_{\theta}=300$ to 23,400). 

\subsection{Simulation datasets}
DNS and LES datasets are gathered or re-generated from cases in four prior studies: \cite{pargal2022adverse}, \cite{wu2019effects}, \cite{na1998structure} and \cite{wu2018effects}. The first two are DNS cases. Data are collected directly from simulations of turbulent boundary layer on a flat-plate and that on  a controlled-diffusion (CD) airfoil with matched non-equilibrium APG distributions along the streamwise direction, as discussed in \cite{pargal2022adverse}. The other two studies are DNS and LES studies, respectively, of flat-plate boundary layers with suction and blowing freestream velocities which lead to boundary layer separation and then reattachment; these two cases are rerun to collect velocity and wall-pressure statistics. For the case of \cite{wu2018effects}, this work provides new data as the wall pressure was not discussed previously. 

Here, the streamwise, wall-normal and spanwise directions are denoted as   $x$, $y$ and $z$.  $u$, $v$ and $w$ are the velocity components in those directions, $t$ is time, $P$ is the static pressure, $\rho$ is the density and $\nu$ is the kinematic viscosity. 
An instantaneous flow variable $\phi(x,y,z,t)$ is decomposed as $\phi =  \overline{\phi}( x,y) + \phi'(x,y,z,t)$, where $\overline{(\cdot)}$ denotes averaging in $z$ and time.

The case of \citet{wu2019effects} provides DNS data on a boundary layer developing on the pressure side of a CD airfoil. The compressible Navier-Stokes equations are solved with the multi-block structured code HiPSTAR (High Performance Solver for Turbulence and Aeroacoustics Research)~\citep{Sandberg2015}. The spatial discretization involves both a five-point fourth-order central standard-difference scheme with Carpenter boundary stencils in the streamwise and crosswise directions~\citep{Carpenter1999}, and a spectral method using the FFTW3 library in the spanwise direction. The time discretization is achieved by an ultra-low-storage five-step fourth-order Runge-Kutta scheme~\citep{Kennedy1999}. Characteristic-based boundary conditions are  used to prevent spurious reflections at the computational domain boundaries~\citep{Sandberg2006,Jones2008}. Details of the problem formulation are provided by \citet{wu2019effects}.
%\q{Boundary conditions?}

The case of \citet{pargal2022adverse} is an incompressible DNS of a flat-plate turbulent boundary layer to emulate the boundary layer development on the downstream portion of the CD airfoil flow studied by  \citet{wu2019effects}.
A finite difference solver on a staggered grid was used. It employs  second-order, central differences for all spatial derivatives and  second-order  Adams-Bashforth semi-implicit time advancement. 
To match the pressure gradient parameter ($K$) of the airfoil boundary layer, a streamwise pressure gradient is imposed by prescribing a  streamwise-varying $U_\infty(x)$ at the top boundary of the domain. The wall-normal freestream velocity  $V_\infty(x)$ is then obtained based on the conservation of mass. A fully turbulent boundary layer flow at the inlet of the domain is obtained using the recycling/rescaling method  \citep{lundgeneration}. A convective outflow boundary condition \citep{orlanski1976simple} is used at the outlet and periodic boundary conditions are used in the spanwise direction. Similar discretization methods and boundary conditions were used in  \cite{wu2018effects} and  \cite{na1998structure} with slight variations in details.
Simulations of the  flows in these two studies were rerun, based on the methodologies of \citet{pargal2022adverse}. For the LES case of \cite{wu2018effects}, the governing equations were solved for the filtered quantities. The sub-grid stress tensor was modeled using  a dynamic eddy-viscosity model~\citep{GermanoPMC91,Lilly92}, in which the coefficient was adjusted using the Lagrangian-Averaging procedure~\citep{MeneveauLC96}. Boundary layer developments in the rerun simulations will be compared to those reported in the original studies in Section~\ref{sec:Blayer}.

\subsection{Experimental datasets}
DNS and LES simulations are limited to comparatively low Reynolds numbers. Experimental datasets are gathered from the studies of  \cite{hu2016characteristics}, \cite{fritsch2022fluctuating}, and \cite{goody2004empirical}, which provide ZPG or pressure gradient flow data  with $Re_\theta$ of up to 23,400. A brief description of the  experimental setup of each case is given below.
\cite{hu2018empirical} carried out experiments in an open-jet anechoic test section of Acoustic Wind Tunnel Braunschweig (AWB). Adverse and favorable pressure gradients in flat-plate boundary layers were achieved by placing a rotatable NACA 0012 airfoil  above the flat plate. Wall-pressure statistics were measured with sub-miniature pressure transducers and boundary layer velocity profiles were obtained using hot wires. $Re_{\theta}$ was up to 19,000, with   $\beta=-0.9$ to $16$. The study is among the  few experimental studies that measured wall-pressure statistics across very different flows due to the very wide  ranges of pressure gradient and Reynolds number.
Similarly, \cite{fritsch2022fluctuating} carried out experiments in a subsonic wind tunnel  with a NACA 0012 airfoil  installed in the center of the test section. The boundary layer was tripped at the upstream section, to ensure a fully turbulent boundary layer in the test section. Wall-pressure statistics were measured for non-equilibrium  pressure gradients ranging from $\beta$ of $-0.5$ to 0.5, with $Re_{\theta}$ reaching 18,000. 
\cite{goody2000surface} carried out measurements in the boundary-layer tunnel of the Aerospace and Ocean Engineering department of Virginia Tech. The wall-pressure statistics measurement was limited to ZPG flows but data reached Reynolds numbers as high as $Re_{\theta}=23,400$. 
%This is one of the highest Reynolds number cases in the literature for which  WPS measurements were available.
%Data on the development of both the turbulent boundary layer and wall-pressure statistics from the simulation and experimental studies    will be discussed in the next section.

\section{Boundary layer development}\label{sec:Blayer}

In this section, the streamwise developments of pertinent flow and boundary layer variables  are presented for  cases in the database. The goal is to provide insights on the appropriate choice of  scaling variables for WPS modeling in non-equilibrium flows.

\begin{figure}
\begin{center}
\includegraphics[width=.9\linewidth]{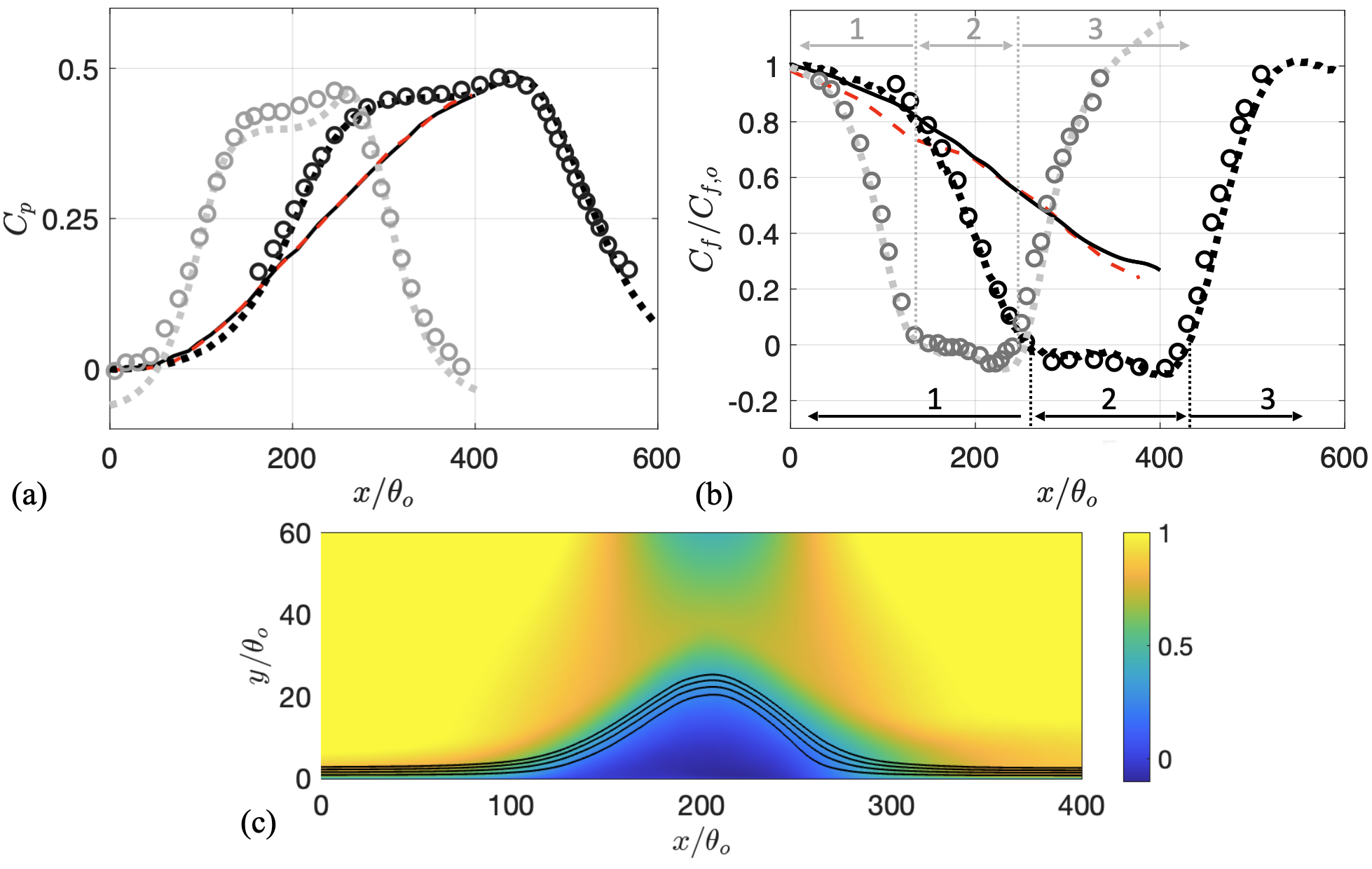}
\end{center}
\caption{Streamwise development of pressure coefficient (a) and friction coefficient (b) in the simulated cases: $\solid$  \cite{pargal2022adverse}, {\color{red}$\dashed$} \cite{wu2019effects},  {\color{gray} $\dotted$}  \cite{wu2018effects},  and {\color{black} $\dotted$}   \cite{na1998structure}, compared to original data of \cite{wu2018effects} and   \cite{na1998structure} ($\circ$).  (c) Contour of mean streamwise velocity normalized by $U_o$ in  \cite{wu2018effects} case, with streamlines shown at stream-function values of $\psi_o=0.5$, 1, 1.5 and 2.}
\label{fig:fig01}
\end{figure}

Figure~\ref{fig:fig01}  shows the variation of the mean wall-pressure coefficient, $C_p=(p|_{y=0}-p_e)/(0.5\,\rho\,U_e^2)$,  and skin friction coefficient, $C_f=(u_\tau/U_e)^2/2$, where  $p_e$ is  edge static pressure at the location of $x=0$, which corresponds to the reference (ZPG) location defined in each of the studies. In Figure~\ref{fig:fig01}(b), the $C_f$ is normalized by its value at the reference location to better compare all cases.
Only simulated cases are presented as the boundary parameters for a continuous range of $x$ are available.
Figures~\ref{fig:fig01}(a) shows that the variations of $C_p$ in the flat-plate \citep{pargal2022adverse} and airfoil \citep{wu2019effects} cases are matched, as required by the flat-plate case setup; comparison between these two flows reveal the effects of convex wall curvature on WPS.
In Figures~\ref{fig:fig01}(b),  the variations of $C_f$ are shown to  match overall for these two cases, with some differences near the airfoil trailing edge due to the convex wall curvature and trailing-edge effects. These comparisons are discussed in  \cite{pargal2022adverse}.
In the cases of \cite{na1998structure} and \cite{wu2018effects} where the flows undergo freestream suction and blowing,  the $C_p$ variation indicates three phases of a separated boundary layer flow (marked in Figures~\ref{fig:fig01}(b)): (1) attached APG flow, (2) separated region and (3) reattached flow under FPG. This is also  reflected in $C_f$ variations: $C_f$ first decreases toward zero in the APG region, reaching negative values in the separated flow region, and increases near the flow reattachment in the FPG region. 
The contour of mean streamwise velocity of the case of \cite{wu2018effects} in Figure~\ref{fig:fig01}(c) confirms these flow stages. 
The coefficients are compared between the results of the present rerun simulations and those of the original studies \citep{na1998structure,wu2018effects}; good match is obtained in both cases.

\begin{figure}
\begin{center}
\includegraphics[width=1.\linewidth]{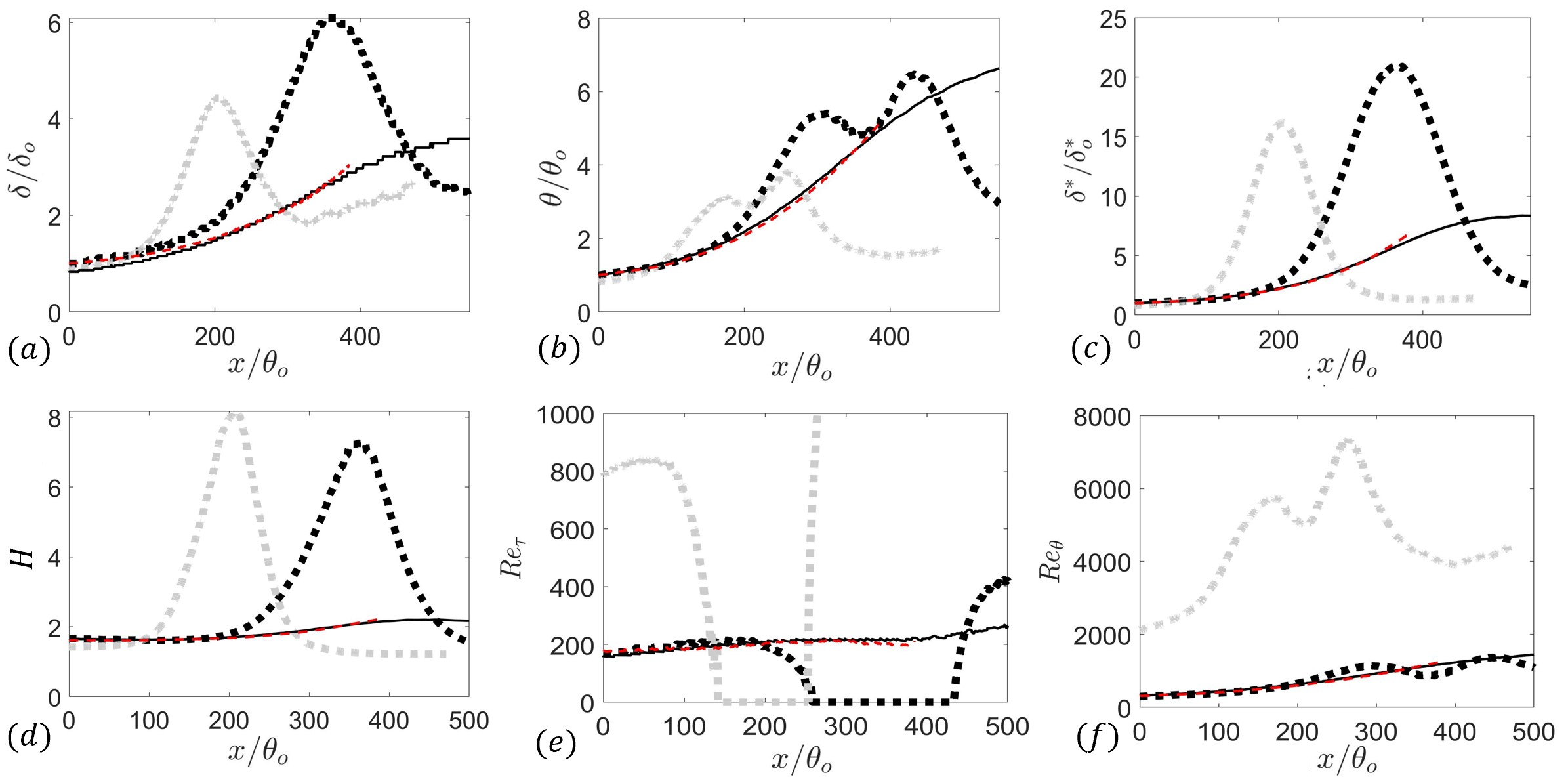}
\end{center}
\caption{Development of boundary layer parameters in the simulated cases. For legend see Figure~\ref{fig:fig01}.
% $\solid$  \cite{pargal2022adverse}, {\color{red}$\dashed$} \cite{wu2019effects}, {\color{gray} $\dotted$} recomputed DNS results of \cite{wu2018effects}, {\color{black} $\dotted$}  recomputed DNS results of  \cite{na1998structure}. 
 }
\label{fig:fig02}
\end{figure}

%Next, i
In Figure~\ref{fig:fig02}, %(a,b,c) 
the variations of boundary layer thickness ($\delta$), momentum thickness ($\theta$) and displacement thickness ($\delta^*$) along the streamwise direction are shown. As expected with an increase in APG, the boundary layer becomes thicker. For the cases with suction and blowing, the thicknesses reach their maxima near the end of the APG zone  and then decrease with FPG.  The development of the shape factor, $H$, in Figure~\ref{fig:fig02}(d)  shows a similar response, which  reflects that the displacement thickness is more sensitive to the pressure gradients  compared to the momentum thickness. 

Reynolds numbers based on different velocity scales are compared. The one based on the inner velocity, $Re_{\tau}=u_\tau \delta/\nu$,   shows a similar trend as that of $C_f$ (Figure~\ref{fig:fig02}(e)), while the one based on edge velocity, $Re_{\theta}=U_e \theta/\nu$, shows a variation similar to that of $\theta$ (Figure~\ref{fig:fig02}(f)). The DNS cases  are conducted in low Reynolds numbers ($Re_\theta\approx 300$ to 1200), while higher values are reached for  the LES case ($Re_\theta\approx 2000$ to 7000).

\begin{figure}
\begin{center}
\includegraphics[width=.8\linewidth]{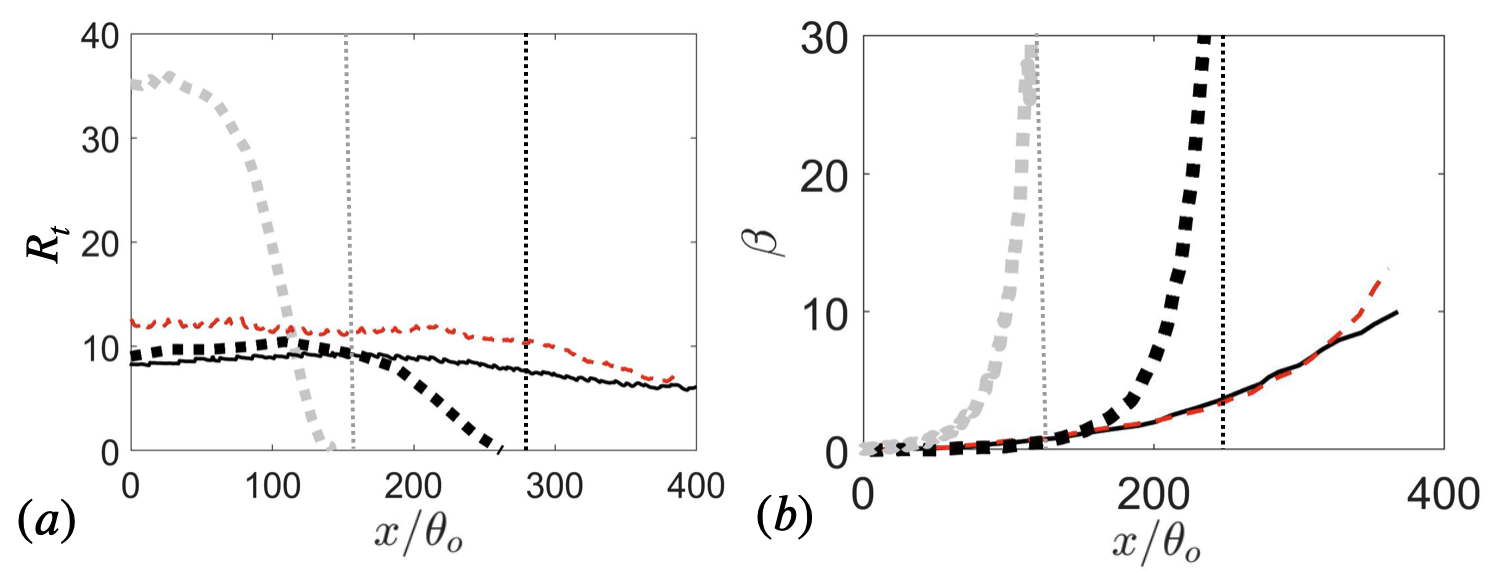}
\end{center}
\caption{Development of $R_t$ and $\beta$ in the simulated cases. For legend see Figure~\ref{fig:fig01}.
%$\solid$  \cite{pargal2022adverse}, {\color{red}$\dashed$} \cite{wu2019effects},  {\color{gray} $\dotted$} recomputed DNS results of \cite{wu2018effects}, recomputed DNS results of {\color{black} $\dotted$} \cite{na1998structure}.
For \citet{wu2018effects} and \cite{na1998structure} cases, only the attached-flow region upstream of the separation point is shown; $\dotted$ (vertical): location of separation point. }
\label{fig:fig03}
\end{figure}

Figure~\ref{fig:fig03} shows the streamwise developments of two boundary layer parameters that are used in most existing WPS models to sensitize the modeled spectrum to the Reynolds number and the pressure gradient: $R_t$
% (\equiv\delta/U_e /\nu/u_{\tau}^2)$
$\equiv Re_\tau(u_\tau/U_e)$ (Figure~\ref{fig:fig03}~(a))
and $\beta$ (Figure~\ref{fig:fig03}~(b)), respectively. 
As the separation point is approached, $R_t$ tends to   0 and $\beta$ to infinity. This indicates issues in many existing WPS models when used for strong-APG flows near incipient separation~\citep{Caiazzo:JFM:2023}, which are examined in detail in Section~\ref{sec:performance_existing}.

\begin{table}
	\begin{center}
		\def~{\hphantom{0}}
		\begin{tabular}{l| cccc} 
			Cases  & $Re_\theta$  & $H$ &$C_f$ & $\beta$ \\ \hline
			         
			\cite{hu2018empirical}, ZPG  & 4889 &  1.41& 0.0025 & 0\\
			
			\cite{hu2018empirical}, APG ($-6^\circ$) & 6979& 1.61& 0.0017 & 3.8\\
			
			\cite{hu2018empirical}, APG ($-10^\circ$) & 8670 & 1.75& 0.0012 & 6 \\
			
			\cite{hu2018empirical}, APG ($-14^\circ$)  & 11046 & 2.12& 0.0006 & 12.5\\
			
			\cite{hu2018empirical}, FPG ($14^\circ$)  & 1940 & 1.26& 0.0068 & -0.5\\ \hline
				
			\cite{fritsch2022fluctuating}, ZPG ($2^\circ$) 
%			(Re=2M, 2 deg.,x=2.47 m.)
			 & 16000 &1.29& 0.0026& -0.02\\
		
		\cite{fritsch2022fluctuating}, APG  ($12^\circ$)  
%		\\ (Re=2M, 12 deg.,x=2.47 m.) 
		 &  18606 & 1.31& 0.0024&  0.58\\
		
			\cite{fritsch2022fluctuating}, FPG ($-10^\circ$)
%			 \\ (Re=2M, -10 deg.,x=2.47 m.)
			   &  14000 & 1.26& 0.0028& -0.47\\ \hline
			\cite{goody2000surface}, ZPG
%			 \\ ZPG (7000) 
			 &7300 & 1.29& 0.0026 & 0\\
			
			\cite{goody2000surface}, ZPG
%			 (23400) 
			 &23400 & 1.29&  0.0022 & 0\\ 
			
			\end{tabular}
			\caption{List of experimental  datasets and values of boundary layer parameters at measurement locations of available data. For \cite{hu2018empirical} and \cite{fritsch2022fluctuating} datasets, %, `deg.' indicates the
   the angle of attack of the airfoil imposed to generate mean pressure gradient is indicated. }

			\label{tab1:exp_param}
		\end{center}
\end{table}

For most of the experimental datasets, streamwise variations of the boundary layer parameters are available at discrete locations only. Representative values of $Re_\theta$, $H$, $C_f$ and $\beta$ are tabulated in  Table~\ref{tab1:exp_param}. 
Specifically, the datasets of \cite{hu2018empirical} experiments contain five cases: one ZPG flow, three APG flows with  $\beta$ varying from 4 to 12, and one FPG flow, at $Re_\theta= 5,000$ to 11,000. The data show that   boundary layer thicknesses (as indicated here by $Re_\theta$; for other thicknesses see the original studies) and the shape factor increase with APG  and decrease in FPG, whereas $C_f$ decreases with  APG and increases in FPG. 
The cases from \cite{fritsch2022fluctuating} are  non-equilibrium  APG and FPG flows but with comparatively milder APG compared to \cite{hu2018empirical}. As a result, the variations in boundary layer parameters are more limited.  Also included are measurements by \citet{goody2000surface}, which were carried out for ZPG flows only but reached higher Reynolds numbers.

%%%%%%%%%%%%%%%%%%%%%%%%%%%%%%%%%%%%%%%%%%%%%%%%%%%%%%%%%%%%%%%%%%%%%%%%%%%%%%%%%%%%%%%%%%%%%%%%%%%%%%%%%%%%%%%%%%%%%%%%%%%%%%%%%%%%%%
\section{Wall-pressure statistics}
\label{sec:p_statistics}

\begin{figure}
\begin{center}
\includegraphics[width=1.\linewidth]{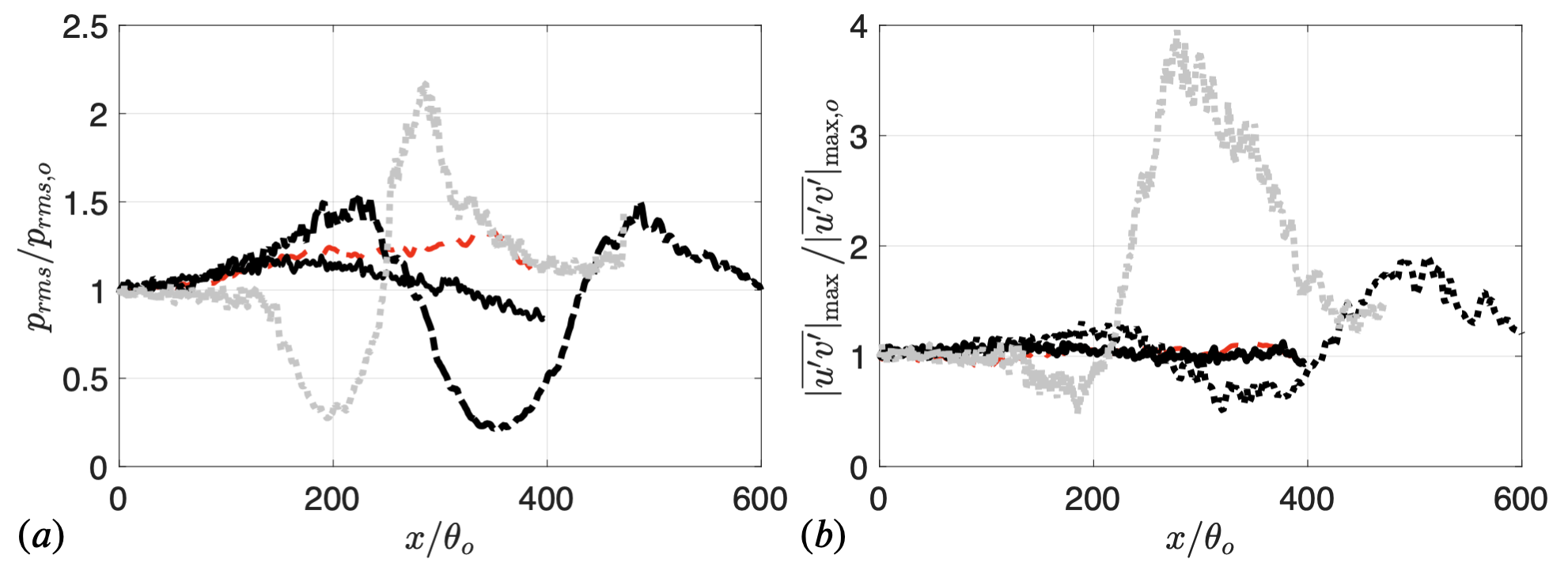}
\end{center}
\caption{(a) Streamwise variation of the wall-pressure r.m.s, normalized by its value at the reference location. (b) Streamwise variation of the local peak magnitude of Reynolds shear stress, normalized by its value at the reference location. For legend see Figure~\ref{fig:fig01}.
%$\solid$ \cite{pargal2022adverse}, {\color{red}$\dashed$} \cite{wu2019effects},  {\color{gray} $\dotted$} recomputed DNS results of \cite{wu2018effects},  {\color{black} $\dotted$} recomputed DNS results of \cite{na1998structure}. 
}
\label{fig:fig04}
\end{figure}

In this section, the development of wall-pressure statistics along  $x$ is discussed using the datasets. Different normalizations are used to analyze the wall-pressure scaling. Effects of  pressure gradient, boundary layer separation and reattachment on the wall-pressure spectrum are  examined. 

For the simulation datasets, streamwise variations of the root-mean-square (r.m.s.) of  wall-pressure fluctuations,  normalized by their specific values at $x=0$,  are compared in Figure~\ref{fig:fig04}(a). 
%Both quantities are normalized by their respective values at the reference location.
%Even for cases with boundary layer separation~\citep{wu2018effects,na1998structure}, the streamwise variation of the r.m.s. wall pressure is less than  50\% for the attached-flow regions. 
Convex wall curvature and trailing edge effect intensify the wall pressure fluctuations, as shown by the comparison between the flat-plate case of \cite{pargal2022adverse} and the airfoil case of \cite{wu2019effects}  for $x/\theta_o>200$.
In the separated-flow region, a drop in wall-pressure fluctuations is seen, which was also observed by \cite{abe2017reynolds}. The dip is attributed to the departure of turbulent eddies from the wall, with mainly large recirculating eddies interacting with the near-wall region.
As the separated shear layer reattaches, the re-emergence  of intense turbulent motions near the wall leads to an augmentation of wall-pressure fluctuations, shown by the $p_{rms}(x)$  maximum shortly after the reattachment point (at $x/\theta_o\approx 280$  for \cite{wu2018effects} and $x/\theta_o\approx 500$ for \cite{na1998structure}). Interestingly, Figure~\ref{fig:fig04}(b) shows that the $x$ variation of the local maximum magnitude of the Reynolds shear stress profile, $\uvmax(x)$,  is very similar to that of $p_{rms}$: the decrease near the separation point and the peak near the flow reattachment occur at almost the same $x$  locations.
Downstream from the reattachment point, as the flow recovers towards the equilibrium ZPG flow, both $p_{rms}$ and  $\uvmax$ reduce towards the ZPG values at $x=0$.

\begin{figure}
\begin{center}
\includegraphics[width=1.\linewidth]{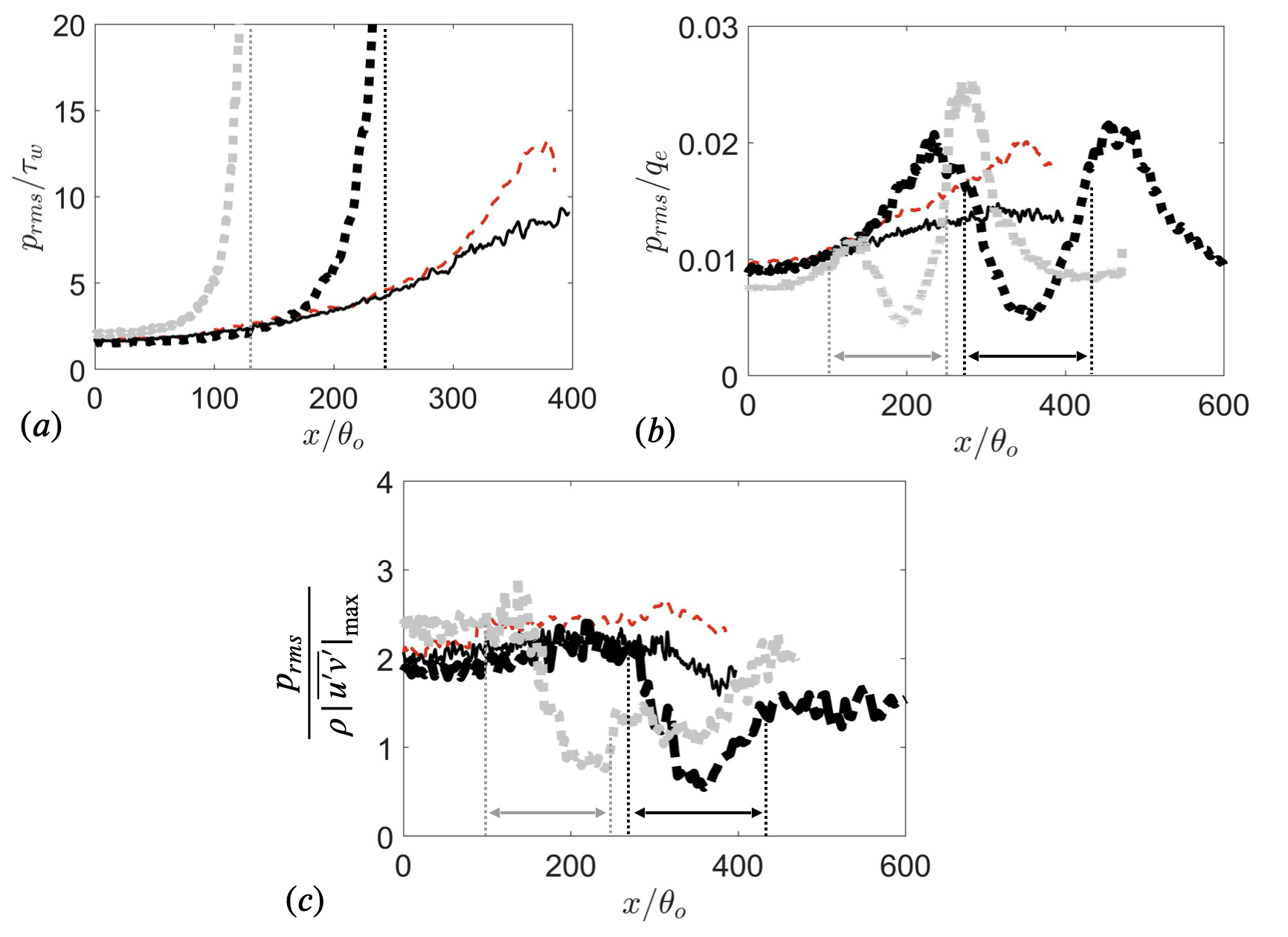}
\end{center}
\caption{ Wall-pressure r.m.s. variation normalized by  (a) local wall shear stress ($\tau_w$), (b) local dynamic pressure ($q_e$), and  (c) local peak magnitude  of Reynolds shear stress. 
%$\solid$  \cite{pargal2022adverse}, {\color{red}$\dashed$} \cite{wu2019effects},  {\color{gray} $\dotted$} recomputed DNS results of \cite{wu2018effects}, recomputed DNS results of {\color{black} $\dotted$} \cite{na1998structure}.
In (a), $\dotted$ (vertical):  locations of separation points. 
In (b) and (c), separated-flow regions are marked. For legend see Figure~\ref{fig:fig01}.
}
\label{fig:fig05}
\end{figure}

In Figure~\ref{fig:fig05} different quantities are used to normalize wall-pressure r.m.s. as it varies along $x$. The r.m.s. normalized by $\tau_w$ (Figure~\ref{fig:fig05}(a)) increases with APG and tends towards infinity as the separating point is approached. %(Figure~\ref{fig:fig05}(a)). 
The use of $\tau_w$ to scale $p_{rms}$ in WPS models is, therefore,  inappropriate for strong-APG boundary layers. 
%Yet, $\tau_w$ is used in most existing WPS models as the wall-pressure scale.  %The 
The r.m.s. normalized by $q_e=0.5\,\rho\,U_e^2$ (Figure~\ref{fig:fig05}(b)) displays  a significant  increase in the APG zone before the flow separation. %(Figure~\ref{fig:fig05}(c)). 
This is because wall-pressure fluctuations are augmented in the APG region, while the edge velocity decreases. In comparison,  $p_{rms}/(\rho\uvmax)$ stays almost constant as long as the boundary layer is attached, regardless of the pressure gradient (Figure~\ref{fig:fig05}(c)).
In the separated flow regions, however, a dip of  $p_{rms}/(\rho\uvmax)$ is observed, caused by a faster damping of $p_{rms}$ inside the recirculation bubble than that of the Reynolds shear stress in the detached shear layer (as shown in Figure~\ref{fig:fig04}).
%, while the peak is due to a more augmented Reynolds shear stress than wall-pressure fluctuations as the boundary layer reattaches. 
These observations indicate  that wall-pressure r.m.s. scales  better with $\rho\uvmax$ than with  $q_e$ or $\tau_w$, in attached  flows under strong pressure gradients; similar observations were made by  \citet{na1998structure}, \citet{abe2017reynolds} and \citet{Caiazzo:JFM:2023}. However, the appropriate wall-pressure r.m.s. scaling for the separated flow region remains yet to be found; but this is out of the scope of the present work.

\begin{table}
	\begin{center}
		\def~{\hphantom{0}}
		\begin{tabular}{l| cccc} 
			Cases  & Legend \\ \hline

                \cite{pargal2022adverse}, $x/\theta_o=0$ , ZPG ($\beta=0$)  & \tikz[baseline=-1ex]\draw[line width=0.5,solid](0,0)--(0.54,0); \\ 
                \cite{pargal2022adverse}, $x/\theta_o=100$, Low APG ($\beta=0.3$)  & \tikz[baseline=-1ex]\draw[line width=1.2,solid](0,0)--(0.54,0);\\ 
                \cite{pargal2022adverse}, $x/\theta_o=290$, APG ($\beta=5$)  & \tikz[baseline=-1ex]\draw[line width=2,solid](0,0)--(0.54,0); \\ \cite{pargal2022adverse}, $x/\theta_o=340$, High APG ($\beta=8$)  & \tikz[baseline=-1ex]\draw[line width=3,solid](0,0)--(0.54,0);\\ \hline
                 \cite{wu2019effects}, $x/\theta_o=0$ , ZPG ($\beta=0.01$)  & \tikz[baseline=-1ex]\draw[line width=0.5,dashed] (0,0)--(0.54,0); \\ 
                \cite{wu2019effects}, $x/\theta_o=100$,  Low APG ($\beta=0.28$)  & \tikz[baseline=-1ex]\draw[line width=1.2,dashed] (0,0)--(0.54,0);  \\ 
                \cite{wu2019effects}, $x/\theta_o=290$, APG ($\beta=4.8$)  & \tikz[baseline=-1ex]\draw[line width=2,dashed](0,0)--(0.54,0); \\
                \cite{wu2019effects}, $x/\theta_o=340$, High APG ($\beta=8.3$) & \tikz[baseline=-1ex]\draw[line width=3,dashed](0,0)--(0.54,0); \\
                \hline
                 \cite{wu2018effects}, $x/\theta_o=50$, Low APG ($\beta=0.9$)  & \tikz[baseline=-1ex]\draw[red,line width=0.5,solid](0,0)--(0.54,0); \\ 
                \cite{wu2018effects}, $x/\theta_o=105$, Very high APG ($\beta=22.8$)   & \tikz[baseline=-1ex]\draw[red,line width=1.2,solid](0,0)--(0.54,0); \\ 
                \cite{wu2018effects}, $x/\theta_o=120$, Very high APG ($\beta=46.28$)   & \tikz[baseline=-1ex]\draw[red,line width=2,solid](0,0)--(0.54,0); \\ 
                \cite{wu2018effects}, $x/\theta_o=130$, Before flow separation ($\beta=171$)   & \tikz[baseline=-1ex]\draw[red,line width=3,solid](0,0)--(0.54,0); \\ 
                \hline
                 \cite{na1998structure}, $x/\theta_o=50$, ZPG ($\beta=0.018$)  &\tikz[baseline=-1ex]\draw[gray,line width=0.5,solid](0,0)--(0.54,0); \\ 
                \cite{na1998structure}, $x/\theta_o=210$, High APG ($\beta=8$)   &\tikz[baseline=-1ex]\draw[gray,line width=1.2,solid](0,0)--(0.54,0); \\ 
                \cite{na1998structure}, $x/\theta_o=230$, Before flow separation ($\beta=162$) &\tikz[baseline=-1ex]\draw[gray,line width=2,solid](0,0)--(0.54,0); \\ \hline
			         
			\cite{hu2018empirical}, ZPG ($\beta=0.1$)  & \tikz[baseline=-1ex]\draw[black,line width=2]  circle (0.08cm); \\

			\cite{hu2018empirical},High APG ($-10^\circ$) ($\beta=6$) & \tikz[baseline=-1ex]\draw[gray,line width=2]  circle (0.08cm); \\
			
			\cite{hu2018empirical}, Very high APG  ($-14^\circ$) ($\beta=12.5$)  & \tikz[baseline=-1ex]\draw[ gray,line width=1]  circle (0.08cm);\\
			
			\cite{hu2018empirical}, FPG ($14^\circ$)  ($\beta=-0.5$) &\tikz[baseline=-1ex]\draw[ cyan,line width=1]  circle (0.08cm); \\ \hline
				
			\cite{fritsch2022fluctuating}, ZPG ($2^\circ$)  ($\beta=0$) 
%			(Re=2M, 2 deg.,x=2.47 m.)
			 & \tikz[baseline=-1ex]\draw[gray,line width=2,dotted] (0,0)--(0.54,0);\\
		
		\cite{fritsch2022fluctuating}, APG  ($12^\circ$)  ($\beta=0.5$) 
%		\\ (Re=2M, 12 deg.,x=2.47 m.) 
		 & \tikz[baseline=-1ex]\draw[black,line width=2,dotted] (0,0)--(0.54,0); \\
		
			\cite{fritsch2022fluctuating}, FPG ($-10^\circ$)  ($\beta=-0.5$) 
%			 \\ (Re=2M, -10 deg.,x=2.47 m.)
			   & \tikz[baseline=-1ex]\draw[cyan,line width=2,dotted] (0,0)--(0.54,0); \\ \hline
			\cite{goody2000surface}, ZPG (7300)  ($\beta=0$) 
%			 \\ ZPG (7000) 
			  & \tikz[baseline=-1ex]\draw[gray,line width=1.5,dash dot] (0,0)--(0.54,0); \\
			
			\cite{goody2000surface}, ZPG (23400 )  ($\beta=0$) 
%			 (23400) 
			 & \tikz[baseline=-1ex]\draw[black,line width=1.5,dash dot] (0,0)--(0.54,0); \\ 
			
			\end{tabular}
			\caption{ Datasets in attached-flow regions (under ZPG or APG) that are considered in analyses of wall-pressure spectrum. }

			\label{tab:legend}
		\end{center}
\end{table}

\begin{figure}
\begin{center}
\includegraphics[width=1.\linewidth]{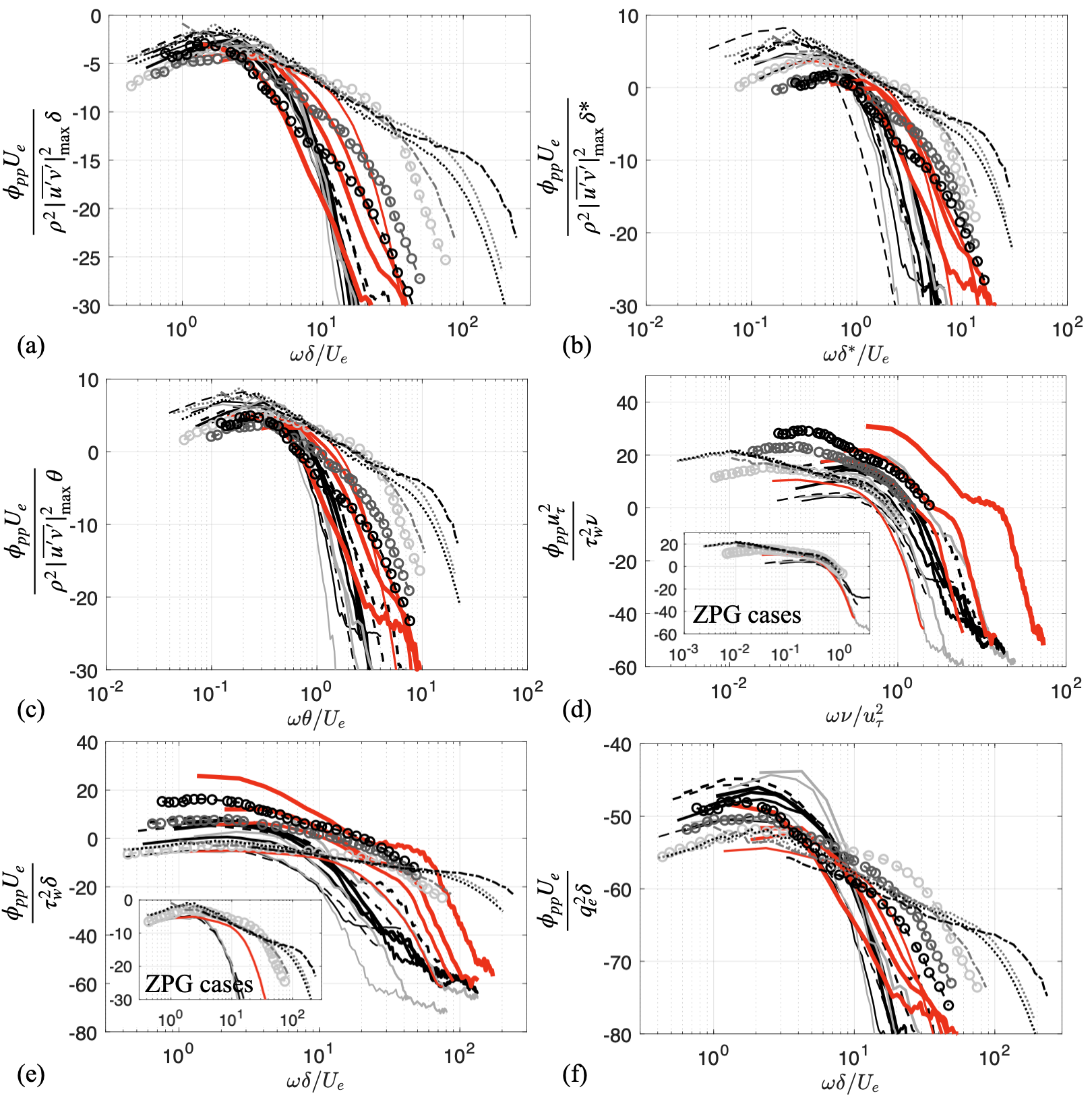}
\end{center}
\caption{Power spectral density (PSD) of wall-pressure fluctuations for attached-flow datasets with ZPG and APGs listed in Table~\ref{tab:legend}, normalized by 
three different scalings involving $\rho\uvmax$ (as the pressure scaling) and outer scales: (a) $\rho\uvmax$, $\delta$ and $U_e$, (b) $\rho\uvmax$, $\delta^*$ and $U_e$ and (c) $\rho\uvmax$, $\theta$ and $U_e$.    
Normalizations with (d) inner scales ($\tau_w$, $\delta_{\nu}$ and $u_\tau$), 
(e) mixed scales ($\tau_w$, $\delta$ and $U_e$), and 
(f) outer scales ($q_e$,  $\delta$ and $U_e$). 
The PSD is evaluated in dB. Insets show ZPG profiles, demonstrating high-frequency collapse in (d) and low-frequency collapse in (e), respectively. 
Legend is listed in Table~\ref{tab:legend}. 
}
\label{fig:fig06}
\end{figure}

The PSD  of the wall-pressure fluctuations (denoted by $\phi_{pp}$) is computed for all simulated and experimental cases  and compared in Figure~\ref{fig:fig06}. Only attached-flow regions, with ZPG or  non-equilibrium APG, are considered here. The $x$ location, $\beta$ value, and legend for each PSD curve  in Figure~\ref{fig:fig06} are listed in Table~\ref{tab:legend}.
Different normalizations are compared. Both $Re_{\theta}$ ($Re_{\theta}=300$ to $23,400$) and $\beta$ ($\beta=0$ to $200$) vary greatly  among these data.  The high $\beta$ values occur near the separation points. 

Figure~\ref{fig:fig06}(a) compares the results  using $-\rho\uvmax$ as the pressure scale, $\delta$  the length scale, and $U_e$ the velocity scale (or equivalently $p_{rms}$, $\delta^*$ and the Zagarola-Smits velocity, as shown by \cite{Caiazzo:JFM:2023}).
Note that in the experimental datasets the Reynolds shear stress data were missing. For these cases, the wall shear stress ($\tau_w$) at a mild-APG ($\beta<1$) location immediately upstream of the APG region, instead of the local Reynolds shear stress, is used to form the pressure scale for the strong-APG region. This approximation is based on the observation that $\uvmax(x)$ does not vary significantly in the attached-flow region upstream of the separation point (Figure~\ref{fig:fig04}(b)). The value of $\uvmax(x)\approx \uvmax(0)$ can then be approximated as $\tau_w(0)$ due to the existence of a constant-stress layer in a boundary layer under zero or mild pressure gradients. The treatment mentioned above is employed for the experimental cases only. 
Under such normalization, Figure~\ref{fig:fig06}(a) shows that approximate low-frequency collapse is obtained. This is expected as  the low-frequency contents are the main contributor to $p_{rms}$, which in turn scales %on 
with $\rho\uvmax$. Swapping the length scale for $\delta^*$ or $\theta$, however, gives more scatter in the low-frequency range (Figures~\ref{fig:fig06}(b) and (c)), as also shown by~\cite{Caiazzo:JFM:2023}. Even though previous works  \citep{kamruzzaman2015semi,abe2017reynolds,Caiazzo:JFM:2023} have shown $\rho\uvmax$ to be the best pressure scaling for wall-pressure spectra, %but 
most of them were limited to low Reynolds number cases with mild pressure gradients. Here, %it is shown that 
the chosen set of scaling  is shown to collapse low-frequency portion of the PSD for a large range of Reynolds number with strong non-equilibrium APG also.

In comparison, normalization based on inner velocity and length scale only (i.e. using $\tau_w$, $\delta_{\nu}\equiv \nu/u_\tau$  and $u_{\tau}$ as shown in Figure~\ref{fig:fig06}(d)) gives a high-frequency collapse for the  ZPG spectra, but a large scatter for the APG ones. When mixed variables are used (i.e. using $\tau_w$, $\delta$  and $U_e$ as shown  in Figure~\ref{fig:fig06}(e)), which is   commonly applied in existing WPS models, the low-frequency range collapses for the ZPG spectra only, but not for cases with strong APG, as $p_{rms}$ does not scale with $\tau_w$. On the other hand, normalization based on outer variables only (i.e. using $q_e$, $\delta$ and $U_e$ as shown in Figure~\ref{fig:fig06}(f)) gives a better collapse than that based purely on the inner variables; however, it still fails to collapse the low-frequency range. Based on these observations, the best $\phi_{pp}$ scaling among these options is thus $(\rho\uvmax)^2\delta /U_e$. 

\begin{figure}
\begin{center}
\includegraphics[width=1.\linewidth]{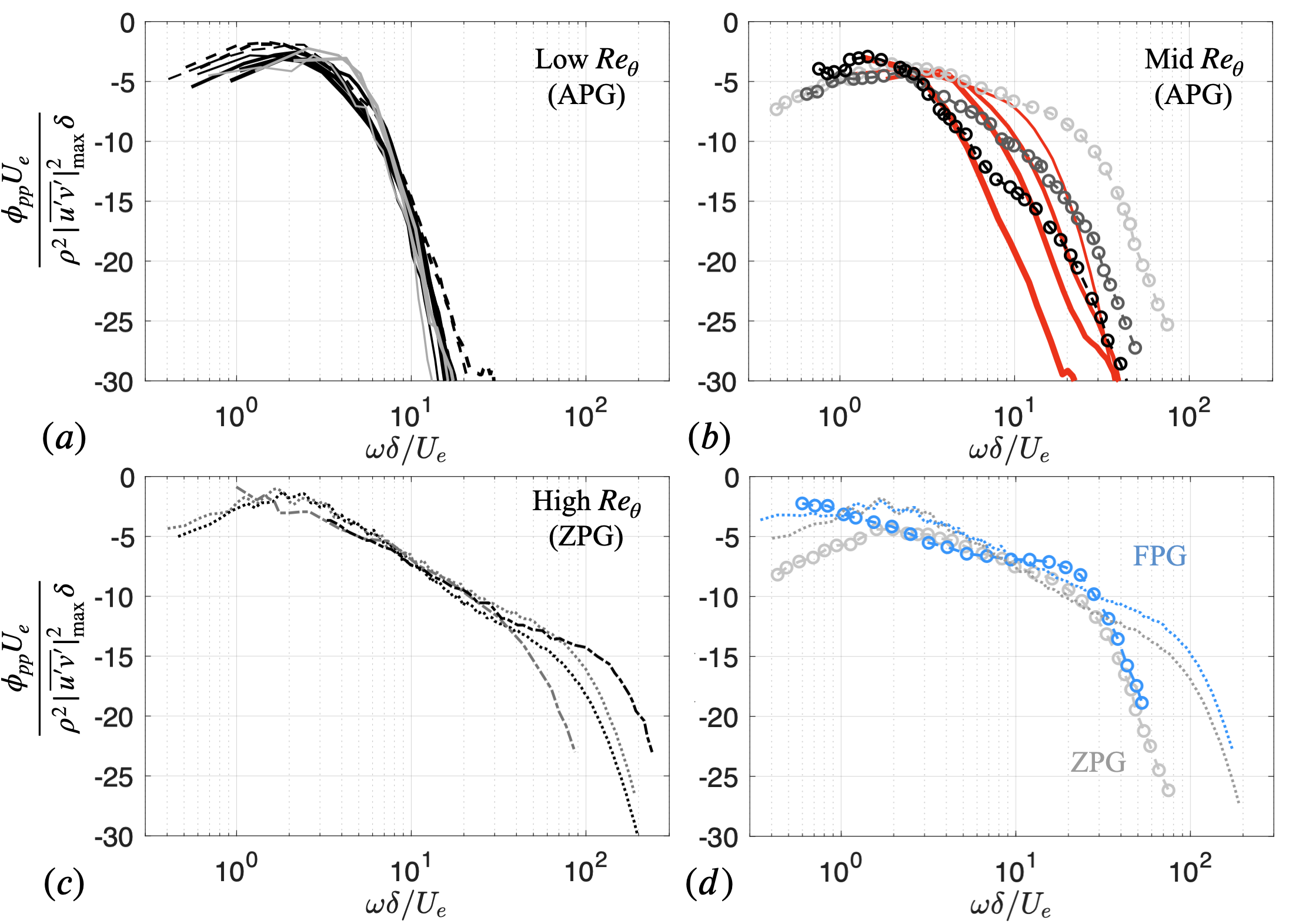}
\end{center}
\caption{ Wall-pressure PSDs in APG boundary layers with different Reynolds number ranges: (a) low-$Re_\theta$ range ($Re_{\theta}=300$ to $1000$),  (b) mid-$Re_\theta$ range ($Re_{\theta}=2000$ to $8000$), and (c) high-$Re_\theta$ range ($Re_{\theta}=8000$ to $23,400$). (d) PSDs in FPG  flows (blue) compared to ZPG ones (gray). See Table~\ref{tab:legend} for legend.}
\label{fig:fig07}
\end{figure}

The effects of Reynolds number are analyzed next. Figures~\ref{fig:fig07}(a-c) categorize the wall pressure PSDs in APG and ZPG flows into three Reynolds number groups: low-$Re_\theta$  ($Re_\theta\approx 300$ to 1000), mid-$Re_\theta$ ($Re_\theta\approx 200$ to 8,000) and high-$Re_\theta$ ($Re_\theta \approx 8,000$ to 23,400) groups. Note that in the high-$Re_\theta$ group, only ZPG or mild-APG flows are available in the present datasets. Figure~\ref{fig:fig07}(a) shows that all low-$Re_\theta$ spectra collapse very well in the majority of frequency range. This is because the overlap range is limited and the low-frequency range is well collapsed by using $\rho\uvmax$ as the pressure scaling.
%due to the use of $\uvmax$ as pressure scaling. 
At higher Reynolds numbers, the overlap range appears and grows with $Re_\theta$ (Figures~\ref{fig:fig07}(b,c)). The width of the overlap range is shown to decrease with APG and the slope of this range increases with APG.

The PSDs in FPG flows are shown in Figure~\ref{fig:fig07}(d) using the two FPG datasets (in blue) of  \cite{hu2018empirical} and  \cite{fritsch2022fluctuating}, as compared to the corresponding ZPG spectra (in gray) from these two studies. Under $\beta\approx -0.5$ (relatively mild FPG), both spectra show a milder slope in the overlap range than the ZPG PSDs. This is consistent with the increase in slope for APG flows discussed above. In addition, the overlap ranges are slightly widened under FPG with the  low-frequency limit moving towards lower frequencies, especially for the lower-Reynolds-number case \citep{hu2018empirical}. This is associated with a weaker mean-flow wake region under FPG.

\begin{figure}
\begin{center}
\includegraphics[width=1\linewidth]{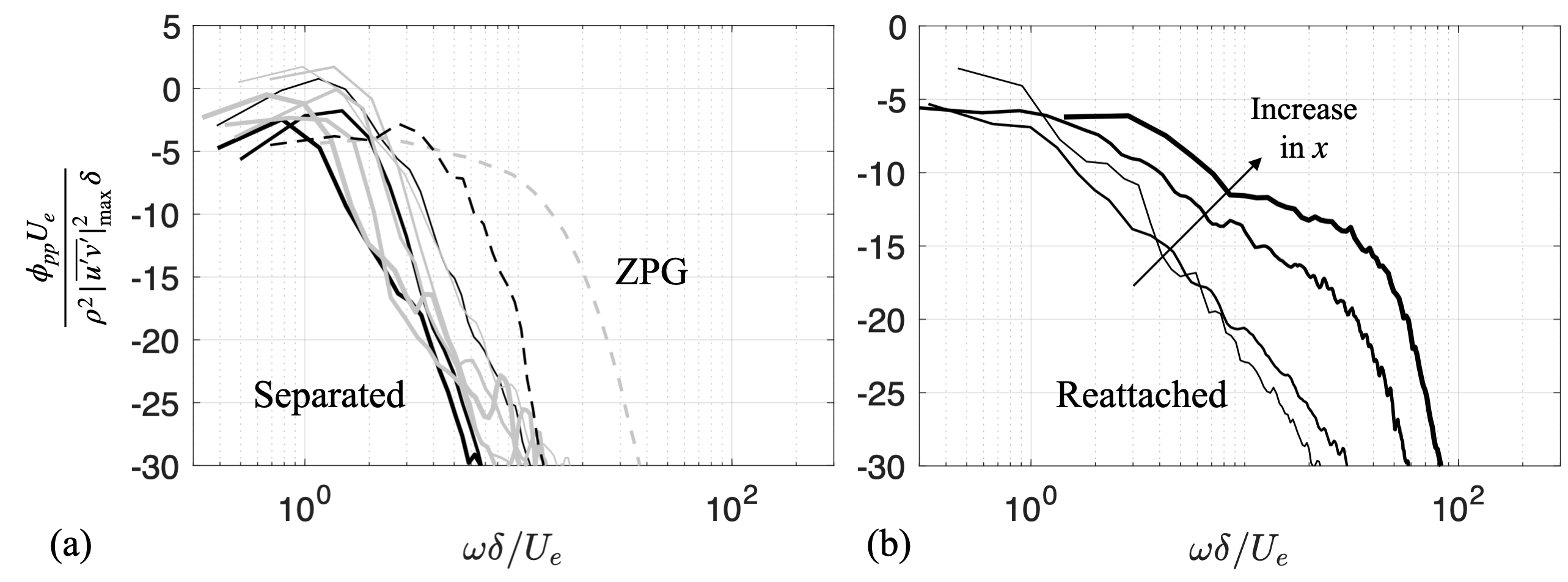}
\end{center}
\caption{(a) Wall-pressure PSDs  in separated-flow regions ($\solid$) compared to those at respective reference ZPG locations  ($\dashed$); gray lines  show \cite{wu2018effects} data at $x/\theta_o = 150$, 175, 200, 220 and 240; black lines show  \cite{na1998structure} data at $x/\theta_o=270$, 300 and 400. 
(b) Wall-pressure PSDs in reattached-flow region of  \cite{wu2018effects}, at $x/\theta_o = 275$, 300, 350 and 450.
Increase in line thickness indicates downstream direction.
%Development of $C_f(x)$ for all DNS datasets, with marked $x$ locations  (vertical lines) in separated-flow regions of cases of \cite{wu2018effects} ({\color{red} $\dashed$}) and \cite{na1998structure} ({\color{blue} $\dashed$}), at which wall-pressure PSDs are compared.  
  }
\label{fig:fig08}
\end{figure}

%\begin{figure}
%\begin{center}
%\includegraphics[width=0.9\linewidth]{./Fig/Fig10_WPS_Reattach}
%\end{center}
%\caption{(a) \q{keep gray curve only; in (b) change m to k.} Development of $C_f(x)$ for all DNS datasets, with marked $x$ locations  (vertical lines) in reattached-flow regions of the case of  \cite{wu2018effects}, at which wall-pressure PSDs are compared.  (b) PSDs of wall-pressure fluctuations  in the $x$ locations marked in (a).  Increase in line thickness indicates downstream direction. }
%\label{fig:fig09}
%\end{figure}

To analyze the WPS associated with separated  and reattached flows, Figures~\ref{fig:fig08}(a) and \ref{fig:fig08}(b) compare the spectra extracted, respectively, from the separated-flow regions and the regions  downstream of the boundary layer reattachment in the cases of \cite{na1998structure}  and \cite{wu2018effects}. 
%The $x$ location of each PSD curve is marked in the $C_f/C_{fo}$ plots (Figures~\ref{fig:fig08}(a) and \ref{fig:fig08(a), respectively).
Figures~\ref{fig:fig08}(a) shows that in the separated-flow regions both overlap-range and high-frequency wall-pressure fluctuations  are  reduced compared to those in respective reference ZPG locations (dashed lines), due to the departure of intense turbulent motions from the wall following the detachment of the shear layer. 
The scaling does not collapse the low-frequency range as it does for attached flows. This is expected as the wall pressure r.m.s  does not scale with $\uvmax$ in this region (Figures~\ref{fig:fig05}(c)). However, it is interesting that the shape of the spectrum does not vary significantly in the separated-flow region: the spectra in  Figures~\ref{fig:fig08}(a) all display a narrow low-frequency peak with greatly reduced high-frequency contribution.
% compared to the PSDs at the reference ZPG locations. 
Downstream from the reattachment point, Figure~\ref{fig:fig08}(b) shows that the spectrum recovers gradually from the low-frequency-dominant state inside the separated-flow region towards the equilibrium state, with augmented mid- to high-frequency contents.

\section{Wall-pressure spectra modelling}
\label{sec:wps}
\subsection{Performance of existing wall-pressure spectra models}
\label{sec:performance_existing}

\begin{table}
	\begin{center}
		\def~{\hphantom{0}}
		\begin{tabular}{l| cccc} 
			Model  & Legend \\ \hline

               \citet{goody2004empirical}  & \tikz[baseline=-1ex]\draw[line width=2,solid](0,0)--(0.54,0);\\ 
                \citet{lee2018empirical}  & \tikz[baseline=-1ex]\draw[line width=2,dashed](0,0)--(0.54,0);\\ 
               \citet{rozenberg2012wall}  & \tikz[baseline=-1ex]\draw[line width=2,dash dot](0,0)--(0.54,0);\\ 
               \citet{hu2018empirical}  & \tikz[baseline=-1ex]\draw[gray,line width=2,solid](0,0)--(0.54,0);\\ 
                \citet{kamruzzaman2015semi}  & \tikz[baseline=-1ex]\draw[gray,line width=2,dashed](0,0)--(0.54,0);\\ 
                Proposed model   & \tikz[baseline=-1ex]\draw[red,line width=2,solid](0,0)--(0.54,0);\\ 
			
			\end{tabular}
			\caption{List of WPS models examined with the present datasets. }

			\label{tab3:legend}
		\end{center}
\end{table}

\begin{figure}
\begin{center}
\includegraphics[width=1\linewidth]{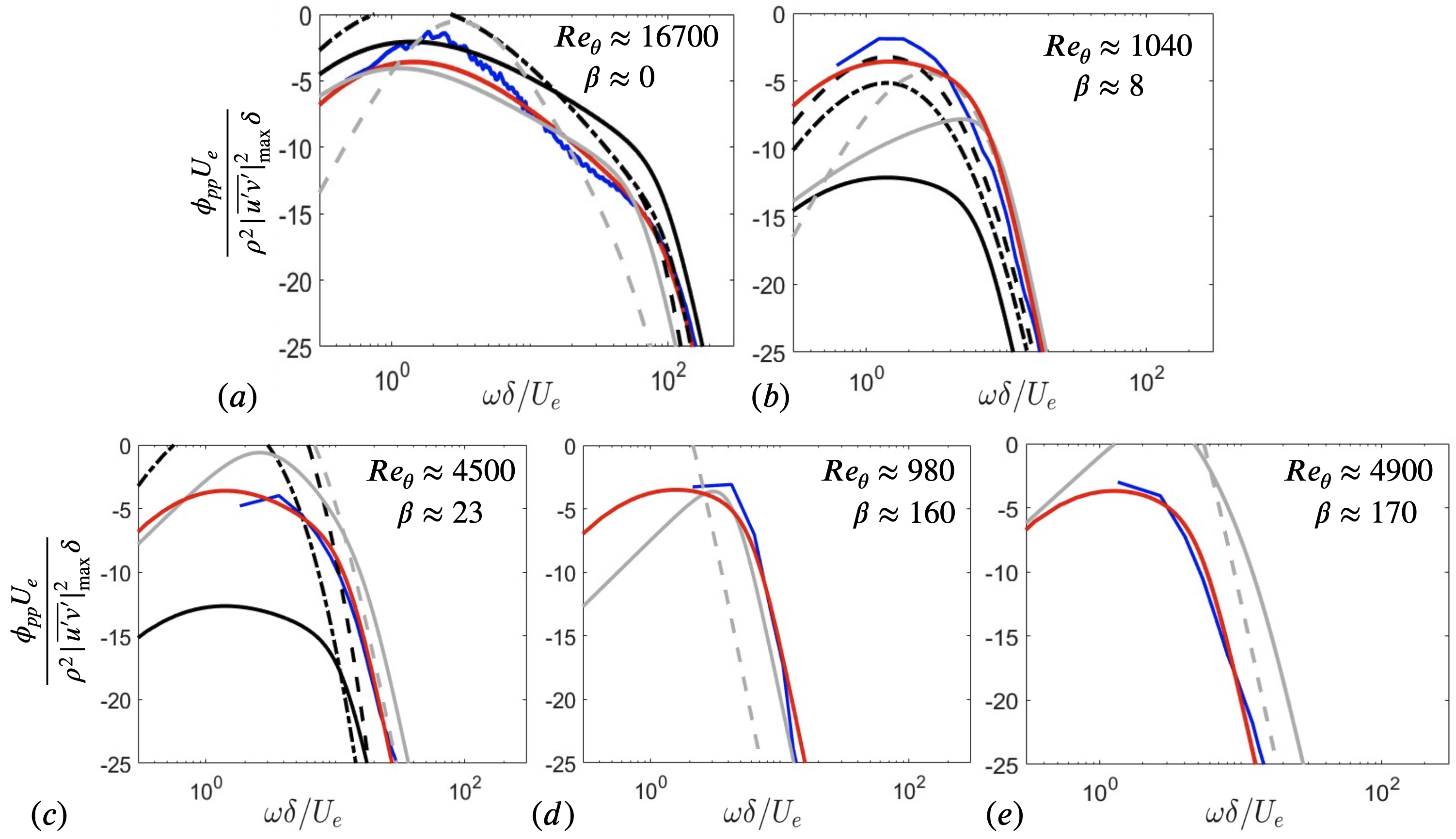}
\end{center}
\caption{Comparison between predictions of WPS models and numerical or experimental measurements ({\color{blue}$\solid$}), for different types of flow:  
(a) weak-APG and high-Reynolds-number flow \citep[][$12^\circ$]{fritsch2022fluctuating}, (b) strong-APG and low-Reynolds-number flow \citep[][]{wu2019effects}, (c) very-strong-APG flow at intermediate Reynolds number \citep[][]{wu2018effects}, (d) flow near separation point at a low Reynolds number \citep[][]{na1998structure}, and  (e) flow near separation point at an intermediate Reynolds number  \citep[][]{wu2018effects}. All models predictions are re-normalized in chosen scaling for comparison purposes. Legend of model results is given in Table~\ref{tab3:legend}.  
 }
\label{fig:fig09}
\end{figure}

Most existing wall-pressure spectral models are developed for regions with zero and adverse pressure gradients. 
Figure~\ref{fig:fig09} compares a number of existing WPS models introduced in Section~\ref{sec:intro} against the present datasets of ZPG and APG (attached regions only) flows (marked by blue solid lines) for a few $Re_\theta$-$\beta$ combinations. Among them, Figures~\ref{fig:fig09}(d,e) show two examples near boundary-layer separation. The models  and their legend are listed in Table~\ref{tab3:legend}. The optimal $\phi_{pp}$ normalization as identified in Section~\ref{sec:p_statistics} is used. Results of the new generalized model (shown by red solid lines) to be introduced in Section~\ref{sec:new_model} is also compared.
The comparison demonstrates dependencies of the WPS on the APG and the Reynolds number as discussed in Section~\ref{sec:p_statistics}. Under the present normalization, the main variations are in the width and slope of the overlap range.

In the (almost) ZPG region (Figure~\ref{fig:fig09} (a)), all  tested models are shown to give reasonably good overall predictions .  \cite{kamruzzaman2015semi} model does not produce an overlap range and, as a result, yields significantly under-predicted high-frequency contents. Moreover, except for %the  
\cite{hu2018empirical} model, all existing models over-predict the spectrum in the overlap range, partially because of a mismatch in the slope. 

For a flow under relatively weak APG ($\beta\approx 8$), Figure~\ref{fig:fig09}(b) shows that \cite{rozenberg2012wall} and \cite{lee2018empirical} models give very good overall predictions, whereas \cite{hu2018empirical} and \cite{kamruzzaman2015semi} models under-predict the WPS at low frequencies. \cite{goody2004empirical} model under-predicts the WPS in the whole frequency range, which  is expected, as this model was developed and calibrated for ZPG flows only. 

In strong-APG flows (Figure~\ref{fig:fig09}(c-e)), especially near the boundary layer separation point,  the existing models  give large errors.
This is  because the model parameters used in these models: $R_t$ and $\beta$, tend towards zero and infinity, respectively. Another source of error is  the inappropriate pressure scaling (i.e. $\tau_w$) used in the  models. For instance,  \cite{rozenberg2012wall} model:  
\begin{equation}
    \label{eq: Rozen}
   \frac{\phi_{pp} (\omega) U_e}{\tau_w^2 \delta^*}=\frac{0.78(1.8\Pi \beta+6)(\omega \delta^*/U_e)^2}{[(\omega \delta^*/U_e)^{0.75}+C_1']^{3.7}+[C_3'(\omega \delta^*/U_e)]^7}, 
\end{equation}
where $C_3'=3.76R_T^{-0.57}$, reaches a singularity  as $\tau_w$ and  $R_t$ become zeros and  $\beta$ becomes infinity. This issue is common in existing models.
In addition, most of these models were fitted to limited types of flows, such as low-Reynolds-number airfoil boundary layers in \cite{kamruzzaman2015semi}, flat-plate boundary layers in \cite{hu2018empirical}, and ZPG flows in \cite{goody2004empirical}. Moreover, sometimes 
the boundary layer flow properties used for model calibration were estimated from lower-fidelity methods such as XFOIL or RANS calculations. Employing $\tau_w$ as the pressure scale for $\phi_{pp}$ also renders the  dimensionless spectrum complicated to model, even for attached flows at strong APGs, as the dimensionless $\phi_{pp}$ displays large variations.
% In summary, the shortcomings of the existing models are: (i) using (sometimes low-fidelity) data from limited types of flows for model development and calibration,  (ii) using wall shear stress as the  pressure scale, and (iii) using $u_\tau$-dependent parameters, such as $R_t$ and $\beta$, to model Reynolds number and pressure gradient effects.
In Section~\ref{sec:new_model}, these limitations are addressed to develop a well-behaved WPS model for both ZPG and APG flows, which can be attached or separated.

\subsection{A new generalized WPS  model for non-equilibrium boundary layers}
\label{sec:new_model}

\citet{goody2004empirical} model is modified by replacing the Reynolds number and pressure gradient parameters with new model inputs that derive directly from the local mean velocity distribution, $U(x,y)$, and are not defined based on $u_\tau$. Only inputs that are quantifiable from engineering predictive approaches (such as RANS models) are considered, so the model is of practical use in engineering applications. The dependence of these inputs on the  $U(y)$ distribution at a given $x$ location allows the model to sense the  local state of the turbulent boundary layer.

\begin{figure}
\begin{center}
\includegraphics[width=1\linewidth]{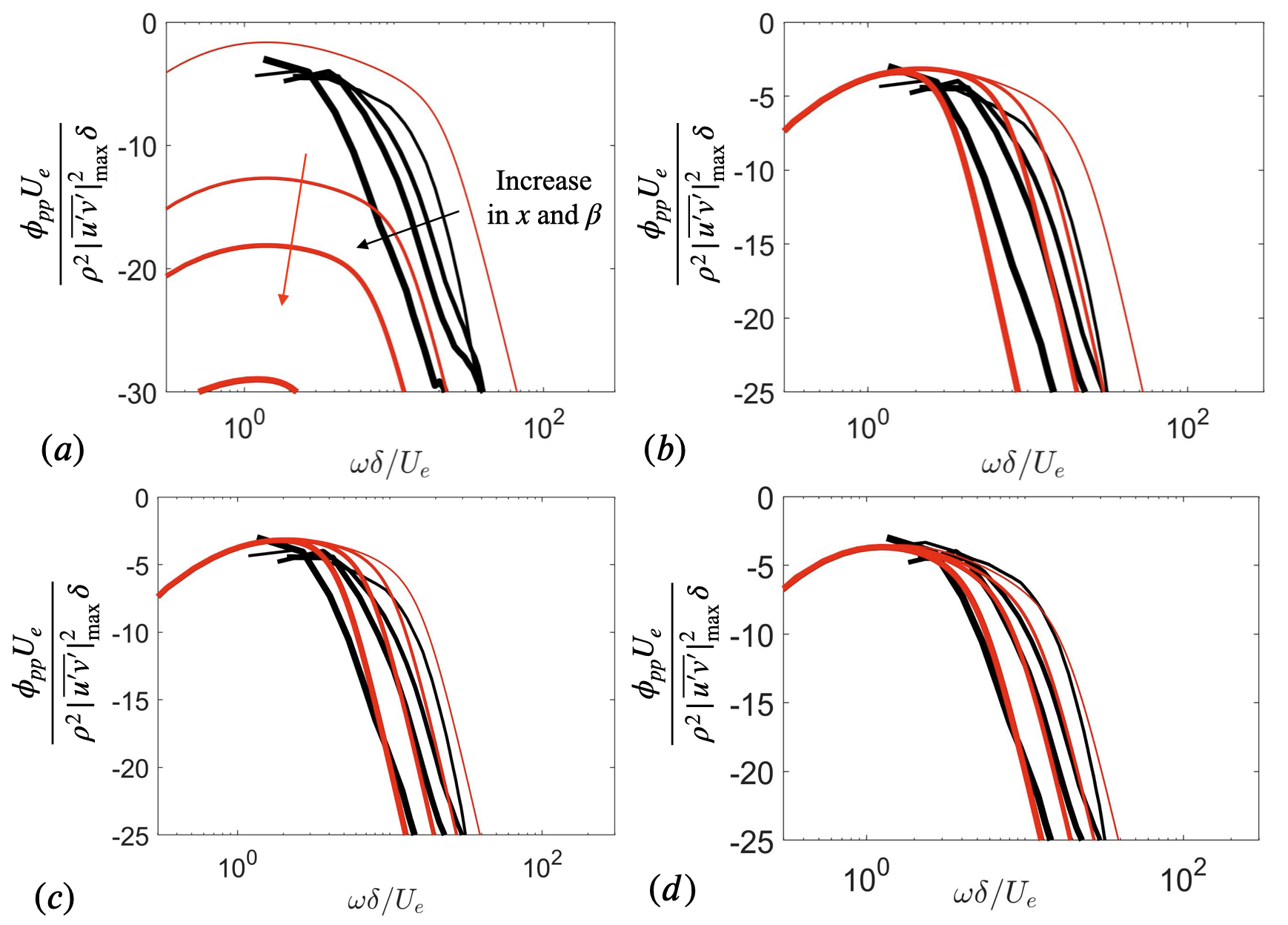}
\end{center}
\caption{Comparison between model predictions of Equation~(\ref{eq:Goody3}) with progressive changes described in text ({\color{red}$\solid$}) and  \cite{wu2018effects} ($\solid$) data in the attached-flow region with weak to strong APGs, to show improvement brought by each model change.
(a) \citet{goody2004empirical} model, (b) pressure scaling changed from $\tau_w$ to $\rho\uvmax$, (c) additionally replacing dependence on $R_t$ with that on $y_w^+$, (d) further addition of  Coles' parameter. Thicker lines indicate increase in $x$ (corresponding to increasing $\beta$).
 }
\label{fig:fig10}
\end{figure}

\citet{goody2004empirical} model reads 
\begin{equation}
    \label{eq:Goody3}
   \widetilde{\phi}_{pp}\left(\widetilde{\omega}\right) =\frac{a\,\widetilde{\omega}^b}{(h\,\widetilde{\omega}^c+d)^{e}+(f \mathcal{F}\,\widetilde{\omega})^g},	
\end{equation}
where $\widetilde{\phi}_{pp}=\phi_{pp}/\phi_{pp}^*$ and $\widetilde{\omega}=\omega/\omega^*$ are dimensionless WPS and frequency, respectively, based on a spectrum scaling of $\phi^*_{pp}$ and a frequency scaling of $\omega^*$ as listed in Table~\ref{t:model_para}. The coefficients $a$ to $h$ and the function $\mathcal{F}$  are defined in Table~\ref{t:model_para}. 
The model was developed based on the observed dependencies of the low-frequency spectrum on the outer scales, as well as that of the  high-frequency spectrum on the inner scales, in a ZPG or weak-pressure-gradient flow.
The model was known to predict well the overlap range of the spectrum for ZPG boundary layers (as also shown in Figure~\ref{fig:fig09}(a)), which depends on the  Reynolds number only. The dependence is captured by including $R_t$ in the model. 
However, the overlap-range spectrum also depends on the pressure gradient in strong-pressure-gradient boundary layers. In Figure~\ref{fig:fig10}(a), the  prediction of the original Goody's model  is examined at several $x$ locations of the \cite{wu2018effects} case in the attached-flow region with weak to strong APGs. In this region,  $\beta$ ranges from 0 to 200 and $Re_{\theta}$ is from 2000 to 6000.
With the increase in APG, Goody's model shows increasing under-prediction in the whole frequency range.

\begin{table}
	\begin{center}
		\def~{\hphantom{0}}
		\begin{tabular}{l| ccccccc} 
                Parameters in Equation~(\ref{eq:Goody3}) & \citet{goody2004empirical} & Present model \\ \hline

			$\phi_{pp}^*$ & $U_e/(\tau_w^2 \delta)$ & $U_e/[(\rho \uvmax)^2 \delta]$\\

            $\omega^*$ & $\delta/U_e$& $\delta/U_e$\\			
			
			$a$ &3 &3\\
			
			$b$ &2 &2\\
			$c$ &0.75 & $\min [1,0.8+(3.34e^{-4})\Pi^{1.86} (y_w^+)^{0.76}]$ \\
			$d$ &0.5 &0.7\\
                $e$ & 3.7& 3.7\\
                $f$ & 1.1 & 1\\
                $g$ & 7 & 7\\
                $h$ & 1 & 1\\
                
            $\mathcal{F}$ 	& $R_t^{-0.57}$ & $(y_w^+)^{-0.37}$ \\

	      \hline

     Separated flow ($\tau_w\le0$) &- & $y_w^+=2$, $c=0.75$, $d=0.5$\\
	 Limited-log-layer flow ($y_w^+<15$) & - & $y_w^+ = 15$, $c=0.85$\\
%	      \hline
			
			\end{tabular}
			\caption{Comparison between \citet{goody2004empirical} model and the proposed model. $\phi_{pp}^*$ and $\omega^*$ are the chosen scalings for  WPS and  frequency, respectively. Special treatments for flow region  characterized by a separated boundary layer or a limited logarithmic layer are listed.  }
        \label{t:model_para}
		\end{center}
\end{table}

In the following, the main model changes (as listed in Table~\ref{t:model_para}) are introduced and progressively applied to demonstrate the improvement of each change. 
First,  the pressure scale (i.e. $\tau_w$) on the left-hand-side of Equation~(\ref{eq:Goody3}) is replaced with $\rho \uvmax$, with the additional change of replacing the constant  $d=0.5$  with 0.7 for a better low-frequency collapse with current data. The effect of these modifications is shown in Figure~\ref{fig:fig10}(b), where the low-frequency range is shown accurately predicted for all the $x$ locations. The use of  $\rho \uvmax$ ensures that the spectral values are finite near the separation point.

\begin{figure}
\begin{center}
\includegraphics[width=0.5\linewidth]{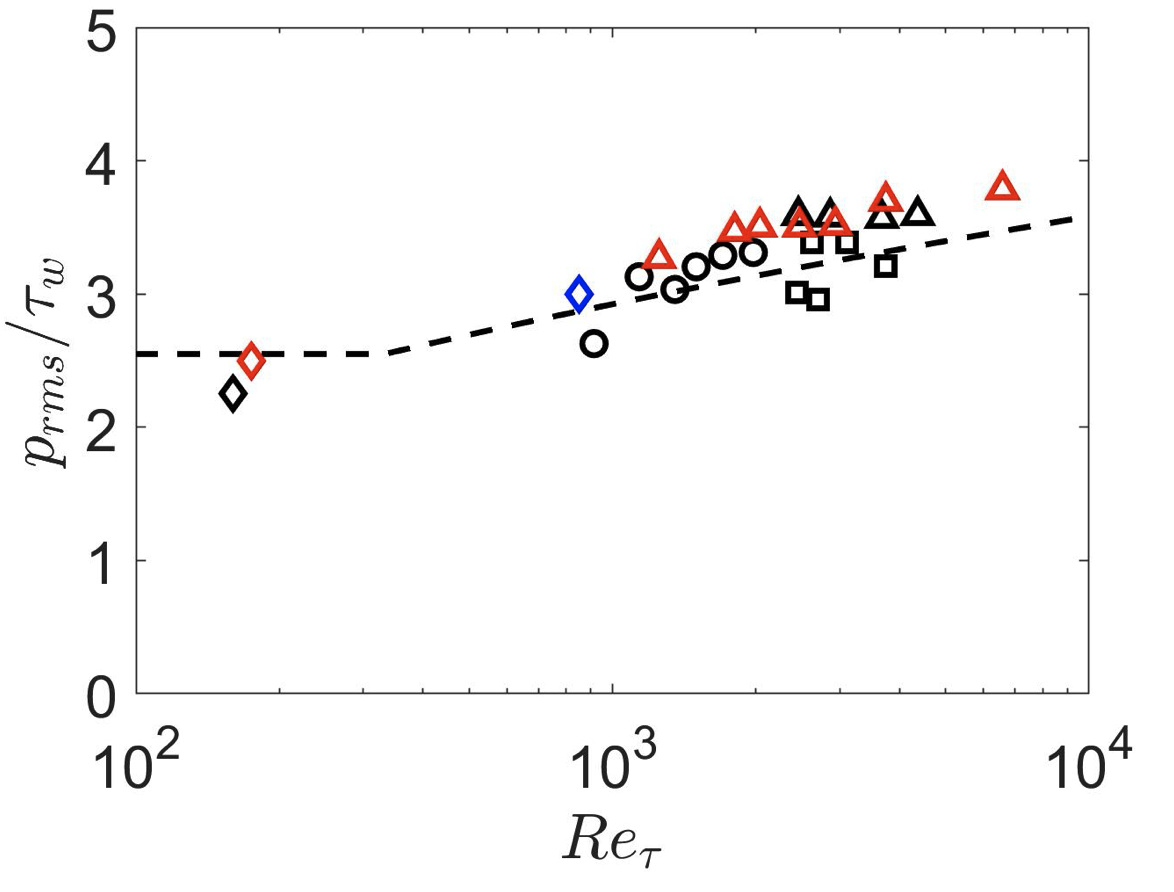}
\end{center}
\caption{ Variation of wall-pressure r.m.s. normalized by $\tau_w$ in ZPG flows: {\color{red} $\Diamond$} \cite{wu2019effects}, {\color{black} $\Diamond$} \cite{pargal2022adverse}, {\color{blue} $\Diamond$} \cite{wu2018effects}, $\triangle$ \cite{blake1970turbulent}, {\color{red} $\triangle$} \cite{simpson1987surface}, $\circ$ \cite{farabee1991spectral} and  $\square$ \cite{bull1976high}. {\color{black}$\dashed$}  $p_{rms}^2/\tau_w^2= 6.5$ for $Re_{\tau}<333$  and  $p_{rms}^2/\tau_w^2= 6.5 + 1.86 \ln(Re_{\tau}/333)$ for  $Re_{\tau}\ge 333$   \citep{farabee1991spectral}. 
%(b) Goody's model predictions with an increase in Reynolds number.
}
\label{fig:fig11}
\end{figure}

Next, to capture the variation of overlap and high-frequency range in APG flows, a new parameter is needed to replace $R_t$ to model the width change of the overlap range. 
Turbulent fluctuations in the logarithmic layer are known as the main contributor to the overlap range of the WPS. For example, \cite{farabee1991spectral} showed that for ZPG boundary layers, with an increase in Reynolds number accompanied by a thickening of the logarithmic layer, the overlap range becomes wider and $p_{rms}/\tau_w$ increases.
This dependency is also shown by the current ZPG datasets in Figure~\ref{fig:fig11}. Additional evidence is provided by \citet{jaiswal2020use}, who 
showed that the logarithmic layer  yields the highest contribution to the overlap range of $\phi_{pp}$ based on analyses of the velocity sources of wall-pressure Poisson's equation.
% evaluated the contribution from different layers of a turbulent boundary layer to $\phi_{pp}$ in various wavenumber ranges, by comparing the velocity sources of the right-hand-side of the wall-pressure  Poisson's  equation  in different layers  based on experimental data. Results (obtained at  $Re_{\tau}=200$)  . 
% In Goody's model, the dependence of the overlap range on the logarithmic layer is modeled by $R_t$. As $R_t$ does not contain pressure-gradient effects, 
Here, a new model input is introduced:  $y_w^+(x)$, defined as the local elevation of the upper edge  of the logarithmic layer, to sensitize the model spectrum to the change in logarithmic layer thickness due to Reynolds number and/or pressure gradients. Specifically,  the term $1.1R_t^{-0.57}$ is replaced with $(y_w^+)^{-0.37}$, with the constant fitted from present data.

\begin{figure}
\begin{center}
\includegraphics[width=1\linewidth]{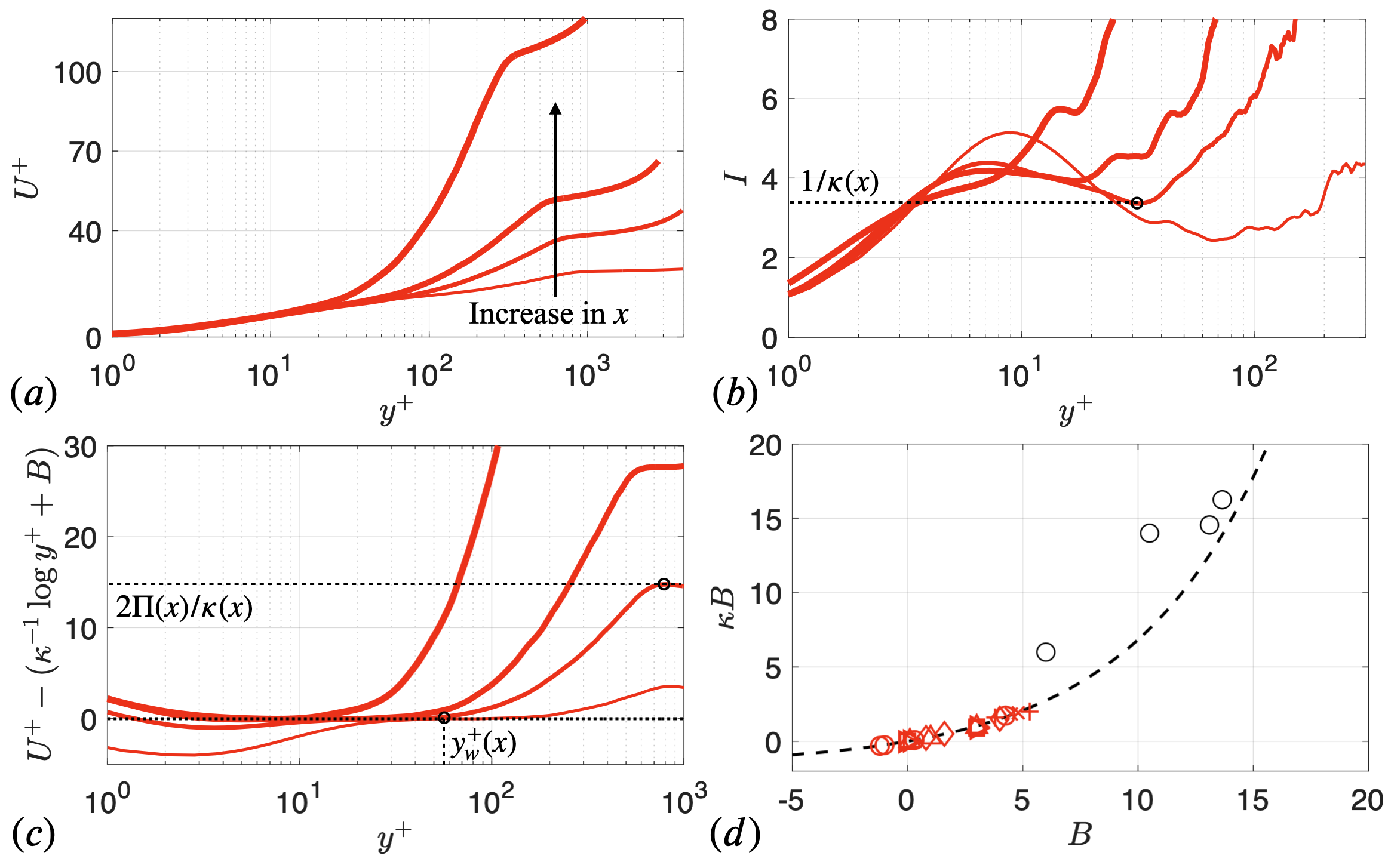}
\end{center}
\caption{(a-c) Calculation of $\kappa(x)$, $B(x)$ and model parameters shown using  \cite{wu2019effects} data at locations $x/\theta_o=50$,  105, 120 and 130. (a) Mean velocity profiles in inner units. Increasing line thickness  indicates increase in $x$. (b) Diagnostic function, $I=y^+\partial U^+/\partial y^+$. (c) Velocity profiles with logarithmic relation subtracted. Calculations of $\kappa$, $\Pi$ and $y_w^+$ are indicated for $x/\theta_o=105$ in (b,c).  (d) Correlation between calculated $\kappa B$ and $B$, compared to  the fitted relation from \cite{nagib2008variations} ($\dashed$): {\color{red} $\circ$} \cite{wu2018effects} (attached flow before separation), {$\circ$} \cite{wu2018effects} (attached flow downstream of reattachment), {\color{red} $\Diamond$} \cite{hu2018empirical}, {\color{red} $\times$} \cite{fritsch2022modeling}, {\color{red} $+$} \cite{goody2004empirical} , {\color{red} $\square$} \cite{pargal2022adverse}, {\color{red} $\rhd$} \cite{wu2019effects} and {\color{red} $\bigtriangleup$} \cite{na1998structure}.  }
\label{fig:fig12}
\end{figure}

The parameter $y_w^+(x)$ is dynamically determined based on the boundary layer mean velocity  $U(x,y)=\overline{u}$, as the $y^+$ location where $U^+(x,y) -[\kappa(x)^{-1} \log y^+ + B(x)]$ departs from $0$ at the upper limit of the logarithmic layer (as shown in Figure~\ref{fig:fig12}(c)). Here, $\kappa$ is the von K\'arm\'an  constant and $B$ is the log-law intercept, both of which allowed to vary along $x$ in a non-equilibrium boundary layer. 
Note that due to the use of local plus units in $y_w^+$ and $U^+$, etc., the present modification needs special treatment at the separation point and inside the separated flow region, which will be discussed later.
To determine  $\kappa(x)$, the diagnostic function $I(x,y)=y^+ \partial U^+/\partial y^+$ (Figure~\ref{fig:fig12}(b)) is calculated from the mean velocity profile (Figure~\ref{fig:fig12}(a)) for data of  \cite{wu2019effects}. The local minima of $I(x,y)$  at a given $x$ is taken as $1/\kappa(x)$. Following the  determination of $\kappa(x)$, $B(x)$ is calculated using $U^+(x,y) -[\kappa(x)^{-1} \log y^+ + B(x)]=0$.  
Figure~\ref{fig:fig12}(d) shows that the correlation between $\kappa(x)$ and $B(x)$ obtained for all cases in the present datasets in the attached flow regions is consistent with   that observed by  \cite{nagib2008variations} and \cite{nickels2004inner} from a large collection of flows with or without pressure gradients. As shown in Figure~\ref{fig:fig10}(c) compared to Figure\ref{fig:fig10}(b), the modeling of overlap-range width based on $y_w^+$ is  successful for the present APG datasets: the high-frequency range is now better predicted with the corrected width. The main WPS prediction error  is now predominantly an inaccurate slope of the overlap range, as shown in Figure~\ref{fig:fig10}(c). 

Recognizing that an APG leads to more energized large turbulence motions in the outer layer and a thinner logarithmic layer, %we hypothesize that 
the change in WPS overlap-range slope (in addition to the change of the width of this range as is characterized by $y_w^+$) is %due to 
assumed to be caused by the variation in the strength of the outer-layer turbulent motions. 
To account for the variation in the strength of the wake region, %we introduce 
an additional model input is used:  Coles' parameter, $\Pi(x)$. An augmentation of $\Pi$  signals stronger turbulent intensity and mixing in the outer layer. Here, $\Pi(x)$ is evaluated based on $U^+(x,y)$, by measuring the peak value of $U^+ -[\kappa^{-1} (\log(y^+) + B)]$  (as shown  in Figure~\ref{fig:fig12}(c))  and dividing it by $2/\kappa(x)$.
In the generic model form in Equation~(\ref{eq:Goody3}), the coefficient $c$ is known to impose the slope of  the overlap range of $\phi_{pp}$ \citep{thomson2022semi,lee2018empirical,rozenberg2012wall}. Therefore, the constant $c$ is replaced by a function of both $\Pi$ and $y_w^+$, fitted based on the present datasets. To keep the model stable, especially in cases with extreme pressure gradient or in the region near flow separation, $c$ is limited to a maximum value of 1.
The final form of the generalized WPS model is
\begin{equation}
    \label{eq:newmodel}
   \frac{\phi_{pp} (\omega) U_e}{(\rho \uvmax)^2 \delta} =\frac{3(\omega \delta/U_e)^2}{[(\omega \delta/U_e)^c+0.7]^{3.7}+[(y_w^+)^{-0.37}(\omega \delta/U_e)]^7}, \quad \text{where}
\end{equation}
\begin{equation}
\label{e:alpha}
	   c = \min\left[1,\;0.8+3.34e^{-4} \Pi^{1.86} (y_w^+)^{0.76}\right].
\end{equation}
Figure~\ref{fig:fig10}(d) shows that the generalized model predicts the WPS very well in both ZPG and strong-APG flows in the present datasets. Note, however, that the present strong-APG data used to calibrate this model are from limited-Reynolds-number flows with a rather narrow WPS overlap range. Additional data from high-Reynolds-number strong-APG flows are not available, but are needed to validate the use of the model of its present form in flows with higher Reynolds numbers.

\begin{figure}
\begin{center}
\includegraphics[width=0.6\linewidth]{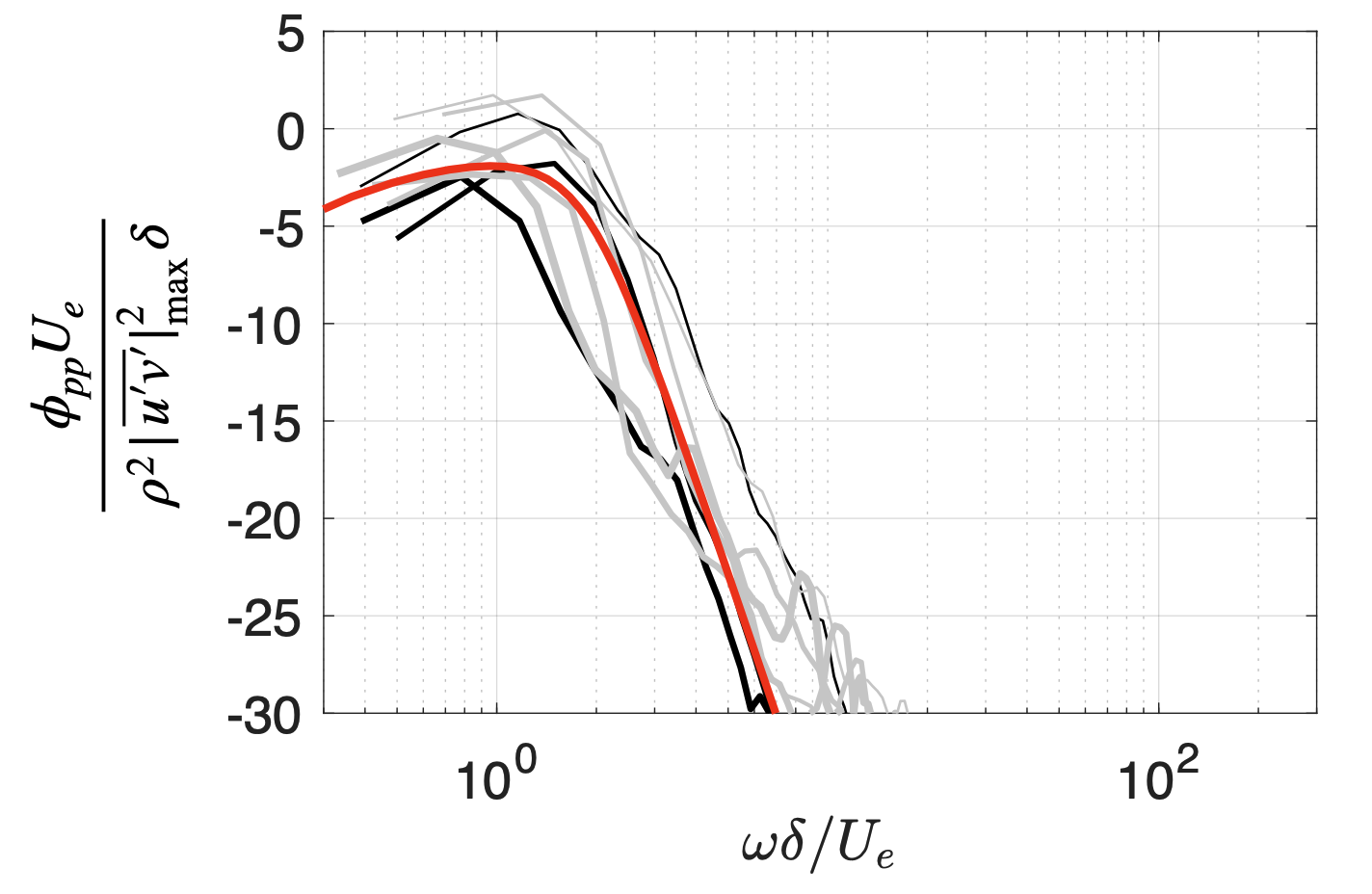}
\end{center}
\caption{Comparison between model prediction ({\color{red} $\solid$}) and simulation data (\solid  \cite{na1998structure} and {\color{gray}{\solid}}  \cite{wu2018effects}) in the separated-flow regions. Increasing line thickness indicates an increase in $x$ as marked in Figure~\ref{fig:fig08}(a).  }
\label{fig:fig13}
\end{figure}

A few  scenarios require special treatments as %follows; the treatments are also
listed in Table~\ref{t:model_para}. For cases with very  low Reynolds numbers with extreme adverse pressure gradient which practically removes the logarithmic region from the boundary layer (i.e. if $y_w^+<15$),  $y_w^+$ and $c$ are set to constant values: $y_w^+=15$ and $c=0.85$ as calibrated from the present datasets, to reflect the insensitivity of $\phi_{pp}$ to either Reynolds number or pressure gradient  as shown in Figure~\ref{fig:fig07}(a).  Moreover, in case of boundary layer separation, modification of the model is needed for $x$ locations at the separation point and inside the separation bubble. Inspired by the almost identical WPS profile across the normalized frequency range in the separated region as shown in Figure~\ref{fig:fig08}(b), 
$y_w^+$ and $c$ are set to constants: $y_w^+=2$, $c=0.75$, and $d$ is set to 0.5 (Table~\ref{t:model_para}), calibrated based on the data of \cite{na1998structure} and  \cite{wu2018effects}. The separation modification is  activated  for $x$ regions where  $C_f(x)$  is calculated as zero or negative, corresponding to the region of mean-flow separation.  
With the above-mentioned  separated-flow treatment, the proposed model is evaluated at a number of streamwise locations inside  the separation regions of the flows of \cite{na1998structure} and  \cite{wu2018effects} in Figure~\ref{fig:fig13}. 
%The examined $x$ locations are marked in Figure~\ref{fig:fig08}(a).
The WPS prediction does not vary with $x$ in this region, since the model parameters are set to constants. 
Good comparison with the simulation data is achieved.
% as the WPS does not vary significantly in separated flows.
Although the separation treatment introduces a discontinuity in $c$ value at the separation point (as $c$ approaches 1 towards the point while $c=0.75$ at that point), it is shown not to affect the prediction significantly, as the overlap range is short in the strong-APG region in the vicinity of the detachment.

\begin{figure}
\begin{center}
\includegraphics[width=1\linewidth]{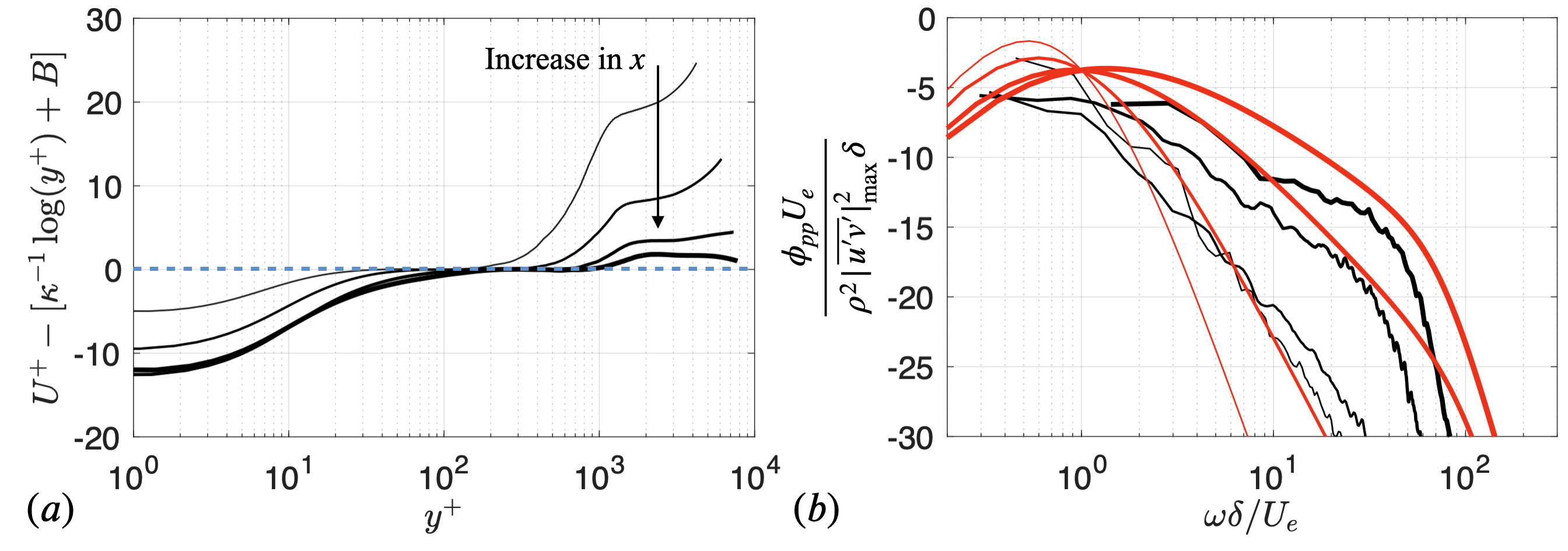}
\end{center}
\caption{Comparison between model prediction ({\color{red} $\solid$}) and   \cite{wu2018effects} data ($\solid$) in the attached-flow region downstream of reattachment point (at $x/\theta_o$ from 250 to 350, the most downstream location corresponding to  near equilibrium flow). (a) Mean velocity profiles with the logarithmic relation subtracted; {\color{blue} $\dashed$} value corresponding to logarithmic layer. 
(b) Wall-pressure PSD comparison. Increasing line thickness indicates an increase in $x$. 
}
\label{fig:fig14}
\end{figure}

Although the model is primarily developed for attached or separated APG flows, it is examined in other regions of a non-equilibrium boundary layer to explore its  extendibility to more universal applications.
First, the model is  evaluated in the region downstream from the flow reattachment point till a near-equilibrium ZPG state in Figure~\ref{fig:fig14} against  \cite{wu2018effects} data, at four $x$ locations between $x/\theta_o=250$ and $x/\theta_o=350$. Figure~\ref{fig:fig14}(a) shows that, near the reattachment point (shown by the thinnest lines),  the local mean velocity  departs significantly from a canonical boundary layer profile, without a clear logarithmic layer. With increasing $x$, the logarithmic layer gradually recovers towards the equilibrium ZPG state and
% , consistent with established understanding, 
the overlap range of the WPS thickens gradually (Figure~\ref{fig:fig14} (b)). The model is shown to  capture  such trend of  WPS variation. The overall spectral levels are well predicted due to the approximate scaling of $p_{rms}$ on $\rho\uvmax$, while the spectral shape is captured by $y_w^+(x)$ and $\Pi(x)$ representing the local thickening of the logarithmic layer and the weakening of wake, respectively, during recovery.

\begin{figure}
\begin{center}
\includegraphics[width=1\linewidth]{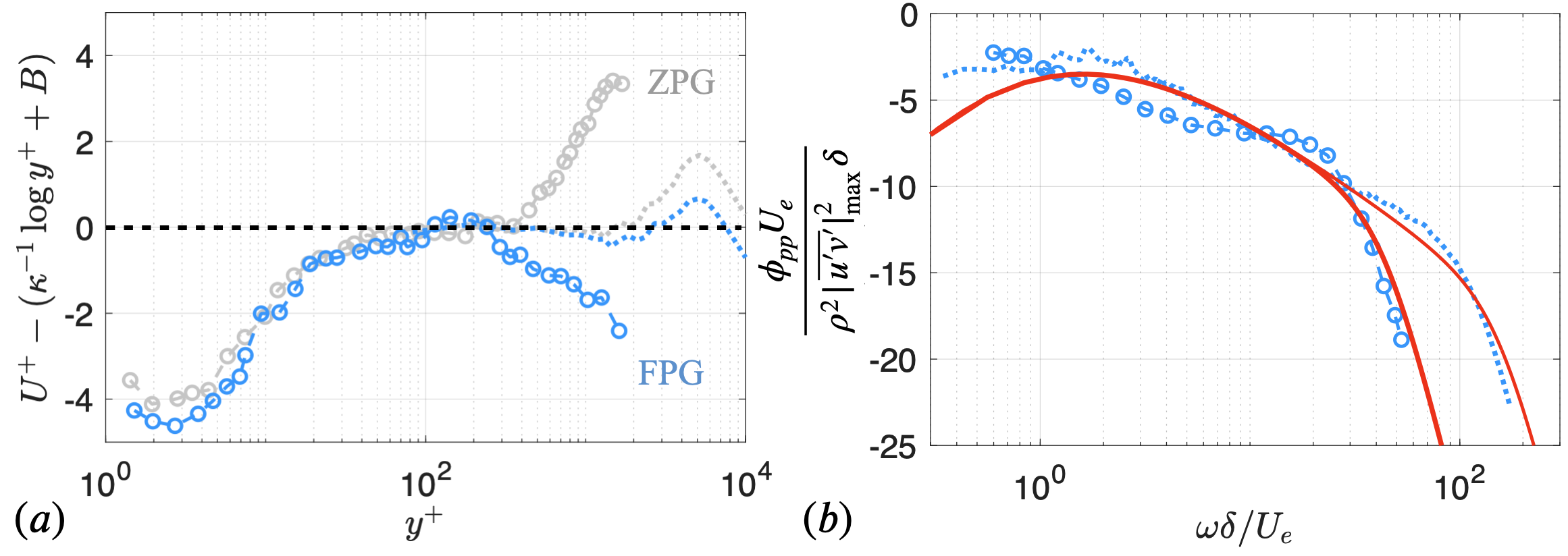}
\end{center}
\caption{ Comparison between model prediction ({\color{red} $\solid$}) and experimental data   in FPG flows \citep{hu2018empirical,fritsch2022fluctuating}. For legend see Table~\ref{tab:legend}.
% ({$\circ$} \cite{hu2018empirical}, $\dotted$ \cite{fritsch2022fluctuating}). 
(a) Mean velocity profile with the logarithmic relation subtracted; FPG and ZPG states are shown in blue and gray, respectively. $\dashed$ Value corresponding to logarithmic layer. (b) Wall-pressure PSD comparison.  
}
\label{fig:fig15}
\end{figure}

In addition,  the model is  tested  in FPG flows as shown in Figure~\ref{fig:fig15}, against the experimental data of \cite{hu2018empirical} and  \cite{fritsch2022fluctuating}. Figure~\ref{fig:fig15}(a) shows that, under  FPGs (shown in blue) as compared to the ZPG profiles  (shown in gray), the main change is a reduction of $\Pi$, which leads to a milder slop of the WPS overlap range. Figure~\ref{fig:fig15}(b) shows that this change in WPS 
%in FPG flows displays an overlap range with a milder slope and a slightly longer width, both of which 
is  globally captured well %overall 
by the proposed model.

\begin{figure}
\begin{center}
\includegraphics[width=1\linewidth]{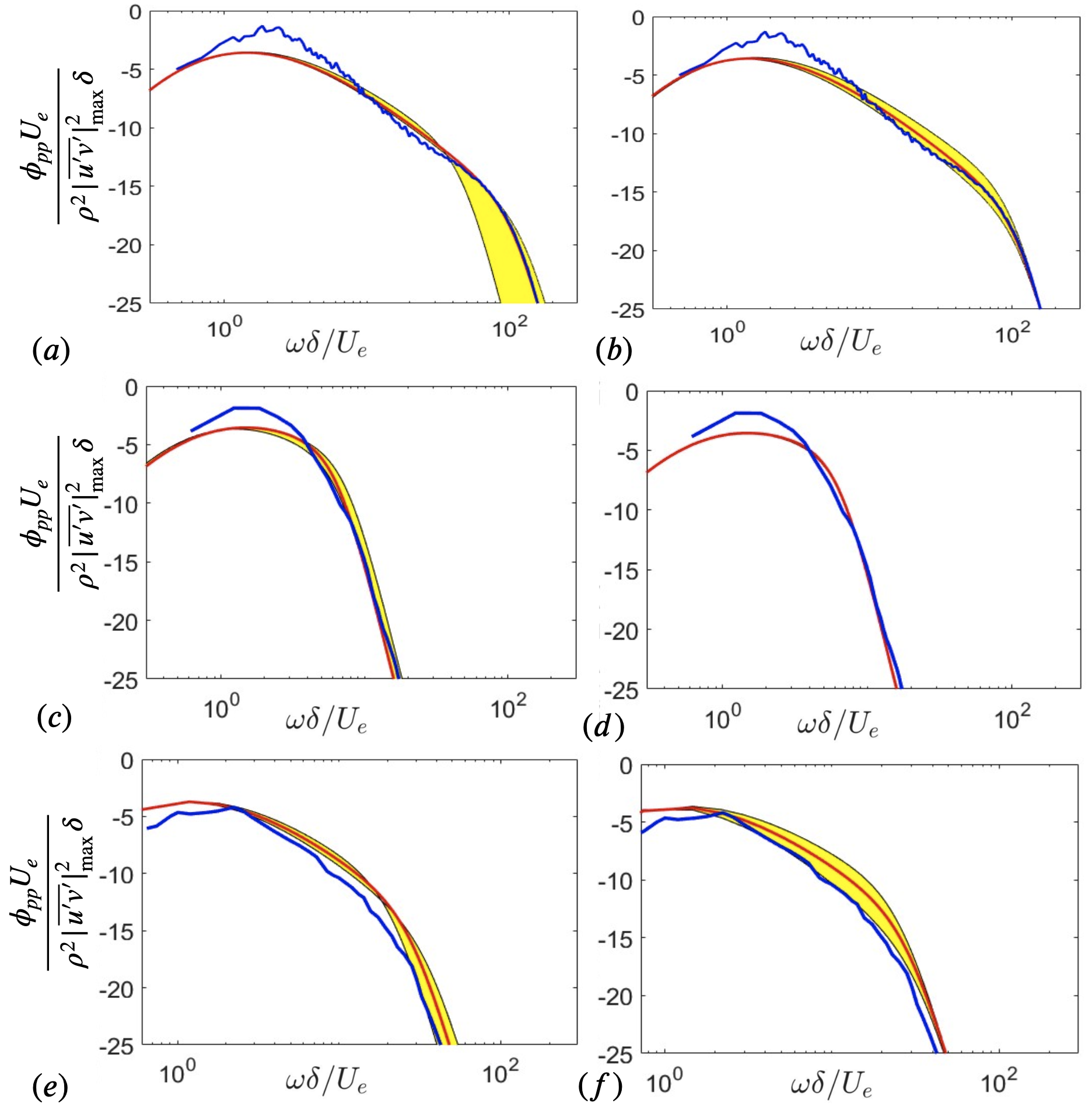}
\end{center}
\caption{WPS prediction of the proposed model ({\color{red}\solid}) compared to measurements ({\color{blue}\solid}) in the following cases: (a,b) high-Reynolds-number and low-APG flow \citep[][$\beta=0.58$]{fritsch2022fluctuating}, (c,d) low-Reynolds-number and strong-APG flow \cite[][$\beta=8.3$]{wu2019effects}, and (e,f) high-Reynolds-number and strong-APG flow  \citep[][$\beta=6$]{hu2018empirical}.  
Yellow regions mark prediction variations with $\pm 30$\% change in input parameters $y_w^+$ (a,c,e)  or $\Pi$ (b,d,f). In (d), variation of $\Pi$ does not change WPS prediction as $c$ is set to a constant at this location due to $y_w^+<15$ (see Table~\ref{t:model_para}).
}
\label{fig:fig16}
\end{figure}

Figure~\ref{fig:fig16} shows results of  sensitivity analyses carried out for the parameters $y_w^+$ and $\Pi$ of  the proposed model. The WPS predictions obtained with $\pm 30$\% change of each of the two parameters (marked by the highlighted region) are compared with the datasets, for three types of flows with different ranges of Reynolds number and $\beta$. 
%For a high-Reynolds-number and weak-APG flow \citep[][$\beta=0.58$]{fritsch2022fluctuating}, 
Figures~\ref{fig:fig16}(a,c,e) show that the variation of $y_w^+$ has an effect on the overlap-range and high-frequency contents, by controlling the width of the overlap range. The effect appears to be particularly strong in a weak-APG flow.
Figures~\ref{fig:fig16}(b,d,f) show that $\Pi(x)$ modifies the slope of the overlap range, with the model particularly sensitive to its value in high-Reynolds-number flows where the overlap range is pronounced. 
These results show that the introduced parameters affect the WPS prediction in their intended ways. Furthermore, slight variations in quantifying $y_w^+$ and $\Pi$ do not significantly worsen WPS prediction.

\section{Conclusions and discussions}
\label{sec:conclusions}

In this study, datasets collected from numerical (DNS and LES) and  experimental studies are used to characterize the variation of wall-pressure statistics in various types of boundary layer flows, attached or separated and then reattached, with  zero, adverse or favorable pressure gradients at different ranges of Reynolds number. The numerical data in the datasets were validated in various flow quantities. They are not prone to errors originating from  installation effects as in many experimental studies and are free from modeling errors as in RANS or boundary layer closures used in the development of some existing WPS models.
% , with flow statistics comparing well against the published literature. The usage of high-fidelity data-sets makes the study less prone to errors originating from either installation effects in experimental studies or low fidelity simulations (XFOIL/ RANS) used in existing WPS model development.
Strongly non-equilibrium streamwise pressure gradient variations are included in the datasets.
By comparing different sets of variables used to normalize the wall-pressure spectrum ($\phi_{pp}$), an optimal set of  scaling is identified: $U_e$, $\delta$ and $\rho\uvmax$, and is used for wall-pressure spectrum model development. 

The performances of various existing wall-pressure spectrum models are evaluated in the flows contained in the datasets. These models are shown to fail to predict the wall-pressure spectra in non-equilibrium strong-APG flows. The failures are caused by the use of inappropriate pressure scaling ($\tau_w$), being fitted to limited types of flows, and the dependencies on $u_\tau$-based model parameters, as $u_\tau$ reduces to zero at the detachment point.

Next, more robust model parameters are proposed and used to modify Goody's model. These parameters are (i) the logarithmic-layer extent, $y_w^+$, and (ii) Coles' parameter, $\Pi$. 
%These two parameters carry information on the local state of the non-equilibrium boundary layer. They replace $R_t$ in the original Goody's model, which serves to gauge indirectly the effects of  pressure gradient and  Reynolds number on the local flow. In addition,  $y_w$ and $\Pi$  contain the `history effects' of pressure gradients as captured in the mean velocity profile.
These parameters carry information on the local state of the boundary layer flow as carried in the mean velocity profile. The parameters are  quantifiable from Reynolds-averaged Navier-Stokes (RANS) calculation of the mean velocity. This is a more direct approach to model the change in contributions of wall-layer and outer-layer turbulent flows to the wall-pressure spectrum, compared to existing approaches based on local pressure gradient (e.g. Clauser's parameter, $\beta$) and/or local Reynolds number (e.g. $R_t$).  
%these parameters determines the averaged condition of turbulent structures using velocity profile.
In addition, $y_w^+$ and $\Pi$ capture  the history effect of non-equilibrium pressure gradients as recorded in the  velocity prediction (from a RANS simulation, for example), which is not directly represented by the local pressure gradient.

Comparison with available numerical and experimental measurements shows that the proposed model gives good predictions for ZPG, APG (attached-flow region) and FPG flows.  For strong APG flows with boundary-layer separation and reattachment, the wall-pressure spectra are shown to display similar shapes and magnitudes across the separation bubble. 
% Hence, are easily modelled into the new model, by fixing parameter values, when $C_f<0$. 
There, the model is shown to give overall good predictions, if the overlap-range width and slope are set to constants fitted based on present data.
%$y_w^+$ and $\Pi$ are kept constants equal to their values at the detachment point. 
A qualitatively good prediction is also obtained downstream of flow reattachment where the boundary layer departs significantly from its equilibrium state.
Hence, the new model  is considered as a generalized wall-pressure spectral model for a wide range of equilibrium and  non-equilibrium boundary layers, as opposed to existing models designed for limited types of flows.

\section*{Acknowledgments} 
SP is grateful for the  funding provided by the Consortium for the Development of Ultra-High Efficiency Quiet Fans at Universit\'e de Sherbrooke. SP and JY also gratefully acknowledge the additional financial support by Office of Naval Research (Award No. N00014-17-1-2102). Computational support was provided by Michigan State University's Institute for Cyber-Enabled Research and the Digital Research Alliance of Canada.

%\backsection[Declaration of interests]{The authors report no conflict of interest.}
\section*{Declaration of interests} The authors report no conflict of interest.

\bibliographystyle{jfm}
\bibliography{jfm_modJY.bib}

\begin{thebibliography}{74}
\expandafter\ifx\csname natexlab\endcsname\relax\def\natexlab#1{#1}\fi
\def\au#1{#1} \def\ed#1{#1} \def\yr#1{#1}\def\at#1{#1}\def\jt#1{\textit{#1}}
  \def\bt#1{#1}\def\bvol#1{\textbf{#1}} \def\vol#1{#1} \def\pg#1{#1}
  \def\publ#1{#1}\def\arxiv#1{#1}\def\org#1{#1}\def\st#1{\textit{#1}}

\bibitem[Abe(2017)]{abe2017reynolds}
{\sc \au{Abe, H.}} \yr{2017}  \at{{R}eynolds-number dependence of wall-pressure
  fluctuations in a pressure-induced turbulent separation bubble}.  \jt{Journal
  of Fluid Mechanics}  \bvol{28}~(4),  \pg{719--754}.

\bibitem[Amiet(1976)]{amiet1976noise}
{\sc \au{Amiet, R.~K.}} \yr{1976}  \at{Noise due to turbulent flow past a
  trailing edge}.  \jt{Journal of Sound and Vibration}  \bvol{47}~(3),
  \pg{387--393}.

\bibitem[Avallone {\em et~al.\/}(2018)Avallone, Van Der~Velden, Ragni \&
  Casalino]{avallone2018noise}
{\sc \au{Avallone, F.}, \au{Van Der~Velden, W. C.~P.}, \au{Ragni, D.} \&
  \au{Casalino, D.}} \yr{2018}  \at{Noise reduction mechanisms of sawtooth and
  combed-sawtooth trailing-edge serrations}.  \jt{Journal of Fluid Mechanics}
  \bvol{848},  \pg{560--591}.

\bibitem[Blake(1970)]{blake1970turbulent}
{\sc \au{Blake, William~K}} \yr{1970}  \at{Turbulent boundary-layer
  wall-pressure fluctuations on smooth and rough walls}.  \jt{Journal of Fluid
  Mechanics}  \bvol{44}~(4),  \pg{637--660}.

\bibitem[Borelli {\em et~al.\/}(2021)Borelli, Gaggero, Rizzuto \&
  Schenone]{borelli2021onboard}
{\sc \au{Borelli, Davide}, \au{Gaggero, Tomaso}, \au{Rizzuto, Enrico} \&
  \au{Schenone, Corrado}} \yr{2021}  \at{Onboard ship noise: Acoustic comfort
  in cabins}.  \jt{Applied Acoustics}  \bvol{177},  \pg{107912}.

\bibitem[Bull(1996)]{bull1996wall}
{\sc \au{Bull, M.~K.}} \yr{1996}  \at{{Wall-pressure fluctuations beneath
  turbulent boundary layers: Some reflections on forty years of research}}.
  \jt{Journal of Sound and Vibration}  \bvol{190}~(3),  \pg{299--315}.

\bibitem[Bull \& Thomas(1976)]{bull1976high}
{\sc \au{Bull, M.~K.} \& \au{Thomas, A. S.~W.}} \yr{1976}  \at{High frequency
  wall-pressure fluctuations in turbulent boundary layers}.  \jt{Physics of
  Fluids}  \bvol{19}~(4),  \pg{597--599}.

\bibitem[Caiazzo {\em et~al.\/}(2023)Caiazzo, Pargal, Wu, Sanjos{\'e}, Yuan \&
  Moreau]{Caiazzo:JFM:2023}
{\sc \au{Caiazzo, A.}, \au{Pargal, S.}, \au{Wu, H.}, \au{Sanjos{\'e}, M.},
  \au{Yuan, J.} \& \au{Moreau, S.}} \yr{2023}  \at{On the effect of adverse
  pressure gradients on wall-pressure statistics in a controlled-diffusion
  aerofoil turbulent boundary layer}.  \jt{Journal of Fluid Mechanics}
  \bvol{960},  \pg{A17}.

\bibitem[Carpenter {\em et~al.\/}(1999)Carpenter, Nordstr{\"o}m \&
  Gottlieb]{Carpenter1999}
{\sc \au{Carpenter, M.~H.}, \au{Nordstr{\"o}m, J.} \& \au{Gottlieb, D.}}
  \yr{1999}  \at{A stable and conservative interface treatment of arbitrary
  spatial accuracy}.  \jt{J. Comp. Phys.}  \bvol{148},  \pg{341--365}.

\bibitem[Casalino {\em et~al.\/}(2021)Casalino, Grande, Romani, Ragni \&
  Avallone]{casalino2021towards}
{\sc \au{Casalino, D}, \au{Grande, E}, \au{Romani, G}, \au{Ragni, D} \&
  \au{Avallone, F}} \yr{2021} {Towards the definition of a benchmark for low
  Reynolds number propeller aeroacoustics}.  \bt{In {\em Journal of Physics:
  Conference Series\/}}, ,  \vol{vol. 1909},  \pg{p. 012013}. IOP Publishing.

\bibitem[Catlett {\em et~al.\/}(2016)Catlett, Anderson, Forest \&
  Stewart]{catlett2016empirical}
{\sc \au{Catlett, M.~R.}, \au{Anderson, J.~M.}, \au{Forest, J.~B.} \&
  \au{Stewart, D.~O.}} \yr{2016}  \at{Empirical modeling of pressure spectra in
  adverse pressure gradient turbulent boundary layers}.  \jt{AIAA Journal}
  \bvol{54}~(2),  \pg{569--587}.

\bibitem[Celik {\em et~al.\/}(2021)Celik, Jamaluddin, Baskaran, Rezgui \&
  Azarpeyvand]{celik2021aeroacoustic}
{\sc \au{Celik, A.}, \au{Jamaluddin, N.~S.}, \au{Baskaran, K.}, \au{Rezgui, D.}
  \& \au{Azarpeyvand, M.}} \yr{2021} Aeroacoustic performance of rotors in
  tandem configuration.  \bt{In {\em AIAA AVIATION 2021 Forum\/}},  \pg{p.
  2282}.

\bibitem[Chang~III {\em et~al.\/}(1999)Chang~III, Piomelli \&
  Blake]{chang1999relationship}
{\sc \au{Chang~III, P.~A.}, \au{Piomelli, U.} \& \au{Blake, W.~K.}} \yr{1999}
  \at{Relationship between wall pressure and velocity-field sources}.
  \jt{Physics of Fluids}  \bvol{11}~(11),  \pg{3434--3448}.

\bibitem[Chase(1980)]{chase1980modeling}
{\sc \au{Chase, D.~M.}} \yr{1980}  \at{Modeling the wavevector-frequency
  spectrum of turbulent boundary layer wall pressure}.  \jt{Journal of Sound
  and Vibration}  \bvol{70}~(1),  \pg{29--67}.

\bibitem[Clauser(1954)]{clauser1954turbulent}
{\sc \au{Clauser, F.~H.}} \yr{1954}  \at{Turbulent boundary layers in adverse
  pressure gradients}.  \jt{Journal of the Aeronautical Sciences}
  \bvol{21}~(2),  \pg{91--108}.

\bibitem[Cohen \& Gloerfelt(2018)]{cohen2018influence}
{\sc \au{Cohen, E.} \& \au{Gloerfelt, X.}} \yr{2018}  \at{Influence of pressure
  gradients on wall pressure beneath a turbulent boundary layer}.  \jt{Journal
  of Fluid Mechanics}  \bvol{838},  \pg{715--758}.

\bibitem[Coles(1956)]{coles1956law}
{\sc \au{Coles, D.}} \yr{1956}  \at{The law of the wake in the turbulent
  boundary layer}.  \jt{Journal of Fluid Mechanics}  \bvol{1}~(2),
  \pg{191--226}.

\bibitem[Corcos(1964)]{corcos1964structure}
{\sc \au{Corcos, G.~M.}} \yr{1964}  \at{The structure of the turbulent pressure
  field in boundary-layer flows}.  \jt{Journal of Fluid Mechanics}
  \bvol{18}~(3),  \pg{353--378}.

\bibitem[Deshmukh {\em et~al.\/}(2019)Deshmukh, Bhattacharya, Jain \&
  Paul]{deshmukh2019wind}
{\sc \au{Deshmukh, S.}, \au{Bhattacharya, S.}, \au{Jain, A.} \& \au{Paul,
  A.~R.}} \yr{2019}  \at{Wind turbine noise and its mitigation techniques: A
  review}.  \jt{Energy Procedia}  \bvol{160},  \pg{633--640}.

\bibitem[Dominique {\em et~al.\/}(2022)Dominique, Van~den Berghe, Schram \&
  Mendez]{dominique2022artificial}
{\sc \au{Dominique, J.}, \au{Van~den Berghe, J.}, \au{Schram, C.} \&
  \au{Mendez, M.~A.}} \yr{2022}  \at{Artificial neural networks modeling of
  wall pressure spectra beneath turbulent boundary layers}.  \jt{Physics of
  Fluids}  \bvol{34}~(3),  \pg{035119}.

\bibitem[Drela(1989)]{DrelaG89}
{\sc \au{Drela, M.}} \yr{1989} {XFOIL}: an analysis and design system for low
  {R}eynolds number airfoils.  \bt{In {\em Low {R}eynolds number
  aerodynamics\/} (ed. \ed{T.~J. Mueller})},  \st{Lecture Notes in
  Engineering},  \vol{vol.~54},  \pg{pp. 1--12}.  \publ{Berlin:
  Springer-Verlag}.

\bibitem[Farabee \& Casarella(1991)]{farabee1991spectral}
{\sc \au{Farabee, T.~M.} \& \au{Casarella, M.~J.}} \yr{1991}  \at{Spectral
  features of wall pressure fluctuations beneath turbulent boundary layers}.
  \jt{Physics of Fluids A: Fluid Dynamics}  \bvol{3}~(10),  \pg{2410--2420}.

\bibitem[Franco {\em et~al.\/}(2020)Franco, Berry, Petrone, De~Rosa, Ciappi \&
  Robin]{franco2020structural}
{\sc \au{Franco, F.}, \au{Berry, A.}, \au{Petrone, G.}, \au{De~Rosa, S.},
  \au{Ciappi, E.} \& \au{Robin, O.}} \yr{2020}  \at{Structural response of
  stiffened plates in similitude under a turbulent boundary layer excitation}.
  \jt{Journal of Fluids and Structures}  \bvol{98},  \pg{103119}.

\bibitem[Fritsch {\em et~al.\/}(2022{\natexlab{{\em a\/}}})Fritsch,
  Vishwanathan, Roy, Todd~Lowe, Devenport, Croaker, Lane, Tkachenko, Pook,
  Shubham {\em et~al.\/}]{fritsch2022modeling}
{\sc \au{Fritsch, D.~J.}, \au{Vishwanathan, V.}, \au{Roy, C.~J.},
  \au{Todd~Lowe, K.}, \au{Devenport, W.~J.}, \au{Croaker, P.}, \au{Lane, G.},
  \au{Tkachenko, O.}, \au{Pook, D.}, \au{Shubham, S.} \& \au{others}}
  \yr{2022{\natexlab{{\em a\/}}}} Modeling the surface pressure spectrum
  beneath turbulent boundary layers in pressure gradients.  \bt{In {\em 28th
  AIAA/CEAS Aeroacoustics 2022 Conference\/}},  \pg{p. 2843}.

\bibitem[Fritsch {\em et~al.\/}(2022{\natexlab{{\em b\/}}})Fritsch,
  Vishwanathan, Todd~Lowe \& Devenport]{fritsch2022fluctuating}
{\sc \au{Fritsch, D.~J.}, \au{Vishwanathan, V.}, \au{Todd~Lowe, K.} \&
  \au{Devenport, W.~J.}} \yr{2022{\natexlab{{\em b\/}}}}  \at{Fluctuating
  pressure beneath smooth wall boundary layers in nonequilibrium pressure
  gradients}.  \jt{AIAA Journal}  \pg{pp. 1--19}.

\bibitem[Germano {\em et~al.\/}(1991)Germano, Piomelli, Moin \&
  Cabot]{GermanoPMC91}
{\sc \au{Germano, M.}, \au{Piomelli, U.}, \au{Moin, P.} \& \au{Cabot, W.~H.}}
  \yr{1991}  \at{A dynamic subgrid-scale eddy viscosity model}.  \jt{Phys.\
  Fluids A}  \bvol{3},  \pg{1760--1765}.

\bibitem[Ghiglino {\em et~al.\/}(2023)Ghiglino, Pullin, Zhou, Abid \&
  Karabasov]{ghiglino2023towards}
{\sc \au{Ghiglino, A.}, \au{Pullin, S.~F.}, \au{Zhou, B.~Y.}, \au{Abid, H.} \&
  \au{Karabasov, S.~A.}} \yr{2023} Towards an adaptive trailing-edge noise
  model using a data-driven approach.  \bt{In {\em AIAA AVIATION 2023
  Forum\/}},  \pg{p. 4371}.

\bibitem[Goody(2004)]{goody2004empirical}
{\sc \au{Goody, M.}} \yr{2004}  \at{Empirical spectral model of surface
  pressure fluctuations}.  \jt{AIAA Journal}  \bvol{42}~(9),  \pg{1788--1794}.

\bibitem[Goody \& Simpson(2000)]{goody2000surface}
{\sc \au{Goody, M.~C.} \& \au{Simpson, R.~L.}} \yr{2000}  \at{Surface pressure
  fluctuations beneath two- and three-dimensional turbulent boundary layers}.
  \jt{AIAA Journal}  \bvol{38}~(10),  \pg{1822--1831}.

\bibitem[Grasso {\em et~al.\/}(2022)Grasso, Roger \&
  Moreau]{grasso2022advances}
{\sc \au{Grasso, G.}, \au{Roger, M.} \& \au{Moreau, S.}} \yr{2022}
  \at{Advances in the prediction of the statistical properties of wall-pressure
  fluctuations under turbulent boundary layers}.  \jt{Fluids}  \bvol{7}~(5),
  \pg{161}.

\bibitem[Hales \& Ayton(2023)]{hales2023adapting}
{\sc \au{Hales, A.} \& \au{Ayton, L.~J.}} \yr{2023} Adapting a trailing-edge
  noise model to an impedance boundary condition.  \bt{In {\em AIAA AVIATION
  2023 Forum\/}},  \pg{p. 3821}.

\bibitem[Hu(2018)]{hu2018empirical}
{\sc \au{Hu, N.}} \yr{2018}  \at{Empirical model of wall pressure spectra in
  adverse pressure gradients}.  \jt{AIAA Journal}  \bvol{56}~(9),
  \pg{3491--3506}.

\bibitem[Hu {\em et~al.\/}(2013)Hu, Buchholz, Herr, Spehr \&
  Haxter]{hu2013contributions}
{\sc \au{Hu, N.}, \au{Buchholz, H.}, \au{Herr, M.}, \au{Spehr, C.} \&
  \au{Haxter, S.}} \yr{2013} Contributions of different aeroacoustic sources to
  aircraft cabin noise.  \bt{In {\em 19th AIAA/CEAS Aeroacoustics
  Conference\/}},  \pg{p. 2030}.

\bibitem[Hu \& Herr(2016)]{hu2016characteristics}
{\sc \au{Hu, N.} \& \au{Herr, M.}} \yr{2016} Characteristics of wall pressure
  fluctuations for a flat plate turbulent boundary layer with pressure
  gradients.  \bt{In {\em 22nd AIAA/CEAS Aeroacoustics Conference, Lyon,
  France\/}},  \pg{pp. 2749--2767}.

\bibitem[Jaiswal {\em et~al.\/}(2020)Jaiswal, Moreau, Avallone, Ragni \&
  Pr{\"o}bsting]{jaiswal2020use}
{\sc \au{Jaiswal, P.}, \au{Moreau, S.}, \au{Avallone, F.}, \au{Ragni, D.} \&
  \au{Pr{\"o}bsting, S.}} \yr{2020}  \at{On the use of two-point velocity
  correlation in wall-pressure models for turbulent flow past a trailing edge
  under adverse pressure gradient}.  \jt{Physics of Fluids}  \bvol{32}~(10),
  \pg{105105}.

\bibitem[Ji \& Wang(2012)]{ji2012surface}
{\sc \au{Ji, M.} \& \au{Wang, M.}} \yr{2012}  \at{Surface pressure fluctuations
  on steps immersed in turbulent boundary layers}.  \jt{Journal of Fluid
  Mechanics}  \bvol{712},  \pg{471--504}.

\bibitem[Jones {\em et~al.\/}(2008)Jones, Sandberg \& Sandham]{Jones2008}
{\sc \au{Jones, L.~E.}, \au{Sandberg, R.~D.} \& \au{Sandham, N.~D.}} \yr{2008}
  \at{Direct numerical simulations of forced and unforced separation bubbles on
  an airfoil at incidence}.  \jt{J. Fluid Mech.}  \bvol{602},  \pg{175--207}.

\bibitem[Kamruzzaman {\em et~al.\/}(2015)Kamruzzaman, Bekiropoulos, Lutz,
  W{\"u}rz \& Kr{\"a}mer]{kamruzzaman2015semi}
{\sc \au{Kamruzzaman, M.}, \au{Bekiropoulos, D.}, \au{Lutz, T.}, \au{W{\"u}rz,
  W.} \& \au{Kr{\"a}mer, E.}} \yr{2015}  \at{A semi-empirical surface pressure
  spectrum model for airfoil trailing-edge noise prediction}.
  \jt{International Journal of Aeroacoustics}  \bvol{14}~(5-6),  \pg{833--882}.

\bibitem[Kennedy {\em et~al.\/}(1999)Kennedy, Carpenter \& Lewis]{Kennedy1999}
{\sc \au{Kennedy, C.~A.}, \au{Carpenter, M.~H.} \& \au{Lewis, R.~M.}} \yr{1999}
   \at{Low-storage, explicit {R}unge-{K}utta schemes for the compressible
  {N}avier-{S}tokes equations}.  \jt{Applied Numerical Mathematics}  \bvol{35},
   \pg{177--219}.

\bibitem[Kraichnan(1956)]{kraichnan1956pressure}
{\sc \au{Kraichnan, R.~H.}} \yr{1956}  \at{Pressure fluctuations in turbulent
  flow over a flat plate}.  \jt{The Journal of the Acoustical Society of
  America}  \bvol{28}~(3),  \pg{378--390}.

\bibitem[Lallier-Daniels {\em et~al.\/}(2021)Lallier-Daniels, Bolduc-Teasdale,
  Rancourt \& Moreau]{LallierDaniels2021}
{\sc \au{Lallier-Daniels, D.}, \au{Bolduc-Teasdale, F.}, \au{Rancourt, D.} \&
  \au{Moreau, S.}} \yr{2021} Fast multi-objective aeroacoustic optimization of
  propeller blades.  \bt{In {\em {Vertical Flight Society's 77th Annual Forum
  \& Technology Display}\/}}.

\bibitem[Lauzon {\em et~al.\/}(2023)Lauzon, Vincent, Pasco, Grondin \&
  Moreau]{lauzon2023aeroacoustics}
{\sc \au{Lauzon, J.-S.}, \au{Vincent, J.}, \au{Pasco, Y.}, \au{Grondin, F.} \&
  \au{Moreau, S.}} \yr{2023} Aeroacoustics of drones.  \bt{In {\em AIAA
  AVIATION 2023 Forum\/}},  \pg{p. 4524}.

\bibitem[Lee(2018)]{lee2018empirical}
{\sc \au{Lee, S.}} \yr{2018}  \at{Empirical wall-pressure spectral modeling for
  zero and adverse pressure gradient flows}.  \jt{AIAA Journal}  \bvol{56}~(5),
   \pg{1818--1829}.

\bibitem[Lee {\em et~al.\/}(2021)Lee, Ayton, Bertagnolio, Moreau, Chong \&
  Joseph]{lee2021turbulent}
{\sc \au{Lee, S.}, \au{Ayton, L.}, \au{Bertagnolio, F.}, \au{Moreau, S.},
  \au{Chong, T.~P.} \& \au{Joseph, P.}} \yr{2021}  \at{{Turbulent boundary
  layer trailing-edge noise: Theory, computation, experiment, and
  application}}.  \jt{Progress in Aerospace Sciences}  \bvol{126},
  \pg{100737}.

\bibitem[Lilly(1992)]{Lilly92}
{\sc \au{Lilly, D.~K.}} \yr{1992}  \at{{A proposed modification of the Germano
  subgrid-scale closure method}}.  \jt{Phys.\ Fluids A}  \bvol{4},
  \pg{633--635}.

\bibitem[Lund {\em et~al.\/}(1998)Lund, Wu \& Squires]{lundgeneration}
{\sc \au{Lund, T.~S.}, \au{Wu, X.} \& \au{Squires, K.~D.}} \yr{1998}
  \at{Generation of turbulent inflow data for spatially-developing boundary
  layer simulations}.  \jt{Journal of Computational Physics}  \bvol{140}~(2),
  \pg{233--258}.

\bibitem[Luo {\em et~al.\/}(2020)Luo, Chu \& Zhang]{luo2020tip}
{\sc \au{Luo, B.}, \au{Chu, W.} \& \au{Zhang, H.}} \yr{2020}  \at{Tip leakage
  flow and aeroacoustics analysis of a low-speed axial fan}.  \jt{Aerospace
  Science and Technology}  \bvol{98},  \pg{105700}.

\bibitem[Meneveau {\em et~al.\/}(1996)Meneveau, Lund \& Cabot]{MeneveauLC96}
{\sc \au{Meneveau, C.}, \au{Lund, T.~S.} \& \au{Cabot, W.~H.}} \yr{1996}
  \at{{A Lagrangian dynamic subgrid-scale model of turbulence}}.  \jt{J. Fluid
  Mech.}  \bvol{319},  \pg{353--385}.

\bibitem[Moreau \& Roger(2009)]{moreau2009back}
{\sc \au{Moreau, S.} \& \au{Roger, M.}} \yr{2009}  \at{{Back-scattering
  correction and further extensions of Amiet's trailing-edge noise model. Part
  II: Application}}.  \jt{Journal of Sound and Vibration}  \bvol{323}~(1-2),
  \pg{397--425}.

\bibitem[Na \& Moin(1998)]{na1998structure}
{\sc \au{Na, Y} \& \au{Moin, Parviz}} \yr{1998}  \at{The structure of
  wall-pressure fluctuations in turbulent boundary layers with adverse pressure
  gradient and separation}.  \jt{Journal of Fluid Mechanics}  \bvol{377},
  \pg{347--373}.

\bibitem[Nagib \& Chauhan(2008)]{nagib2008variations}
{\sc \au{Nagib, H.~M.} \& \au{Chauhan, K.~A.}} \yr{2008}  \at{Variations of von
  {K}{\'a}rm{\'a}n coefficient in canonical flows}.  \jt{Physics of Fluids}
  \bvol{20},  \pg{101518}.

\bibitem[Nickels(2004)]{nickels2004inner}
{\sc \au{Nickels, T.~B.}} \yr{2004}  \at{Inner scaling for wall-bounded flows
  subject to large pressure gradients}.  \jt{Journal of Fluid Mechanics}
  \bvol{521},  \pg{217--239}.

\bibitem[Orlanski(1976)]{orlanski1976simple}
{\sc \au{Orlanski, I}} \yr{1976}  \at{A simple boundary condition for unbounded
  hyperbolic flows}.  \jt{Journal of Computational Physics}  \bvol{21}~(3),
  \pg{251--269}.

\bibitem[Palani {\em et~al.\/}(2023)Palani, Paruchuri, Joseph, Karabasov,
  Markesteijn, Abid, Chong \& Utyuzhnikov]{palani2023modified}
{\sc \au{Palani, S.}, \au{Paruchuri, C.~C.}, \au{Joseph, P.}, \au{Karabasov,
  S.~A.}, \au{Markesteijn, A.}, \au{Abid, H.}, \au{Chong, T.~P.} \&
  \au{Utyuzhnikov, S.}} \yr{2023} {Modified TNO-Blake model for aerofoil
  surface pressure prediction with canopies}.  \bt{In {\em AIAA AVIATION 2023
  Forum\/}},  \pg{p. 3203}.

\bibitem[Panton \& Linebarger(1974)]{panton1974wall}
{\sc \au{Panton, R.~L.} \& \au{Linebarger, J.~H.}} \yr{1974}  \at{Wall pressure
  spectra calculations for equilibrium boundary layers}.  \jt{Journal of Fluid
  Mechanics}  \bvol{65}~(2),  \pg{261--287}.

\bibitem[Pargal {\em et~al.\/}(2023)Pargal, Li \& Li]{pargal2023large}
{\sc \au{Pargal, S.}, \au{Li, W.} \& \au{Li, Y.}} \yr{2023} {Large Eddy
  Simulation of non-equilibrium flows using Lattice Boltzmann method}.  \bt{In
  {\em AIAA SCITECH 2023 Forum\/}},  \pg{p. 2149}.

\bibitem[Pargal {\em et~al.\/}(2022)Pargal, Wu, Yuan \&
  Moreau]{pargal2022adverse}
{\sc \au{Pargal, S.}, \au{Wu, H.}, \au{Yuan, J.} \& \au{Moreau, S.}} \yr{2022}
  \at{Adverse-pressure-gradient turbulent boundary layer on convex wall}.
  \jt{Physics of Fluids}  \bvol{34}~(3),  \pg{035107}.

\bibitem[Roger \& Moreau(2005)]{roger2005back}
{\sc \au{Roger, M.} \& \au{Moreau, S.}} \yr{2005}  \at{{Back-scattering
  correction and further extensions of Amiet's trailing-edge noise model. Part
  1: theory}}.  \jt{Journal of Sound and Vibration}  \bvol{286}~(3),
  \pg{477--506}.

\bibitem[Rossi \& Sagaut(2023)]{rossi2023prediction}
{\sc \au{Rossi, T.} \& \au{Sagaut, P.}} \yr{2023} Prediction of wall-pressure
  spectra for separated/reattached boundary layer flows.  \bt{In {\em AIAA
  AVIATION 2023 Forum\/}},  \pg{p. 3338}.

\bibitem[Rozenberg {\em et~al.\/}(2012)Rozenberg, Robert \&
  Moreau]{rozenberg2012wall}
{\sc \au{Rozenberg, Y.}, \au{Robert, G.} \& \au{Moreau, S.}} \yr{2012}
  \at{Wall-pressure spectral model including the adverse pressure gradient
  effects}.  \jt{AIAA Journal}  \bvol{50}~(10),  \pg{2168--2179}.

\bibitem[Samarasinghe {\em et~al.\/}(2016)Samarasinghe, Zhang \&
  Abhayapala]{samarasinghe2016recent}
{\sc \au{Samarasinghe, P.~N.}, \au{Zhang, W.} \& \au{Abhayapala, T.~D.}}
  \yr{2016}  \at{{Recent advances in active noise control inside automobile
  cabins: Toward quieter cars}}.  \jt{IEEE Signal Processing Magazine}
  \bvol{33}~(6),  \pg{61--73}.

\bibitem[Sandberg(2015)]{Sandberg2015}
{\sc \au{Sandberg, R.~D.}} \yr{2015}  \at{{Compressible-flow DNS with
  application to airfoil noise}}.  \jt{{Flow, Turb. Comb.}}  \bvol{95},
  \pg{211--229}.

\bibitem[Sandberg \& Sandham(2006)]{Sandberg2006}
{\sc \au{Sandberg, R.~D.} \& \au{Sandham, N.~D.}} \yr{2006}  \at{Nonreflecting
  zonal characteristic boundary condition for direct numerical simulation of
  aerodynamic sound}.  \jt{AIAA J.}  \bvol{44},  \pg{402--405}.

\bibitem[Sanjos\'e \& Moreau(2018)]{Sanjose2018}
{\sc \au{Sanjos\'e, M.} \& \au{Moreau, S.}} \yr{2018}  \at{{Fast and accurate
  analytical modeling of broadband noise for a low-speed fan}}.  \jt{The
  Journal of the Acoustical Society of America}  \bvol{143}~(5),
  \pg{3103--3113}.

\bibitem[Schloemer(1967)]{schloemer1967effects}
{\sc \au{Schloemer, H.~H.}} \yr{1967}  \at{Effects of pressure gradients on
  turbulent-boundary-layer wall-pressure fluctuations}.  \jt{The Journal of the
  Acoustical Society of America}  \bvol{42}~(1),  \pg{93--113}.

\bibitem[Shubham {\em et~al.\/}(2023)Shubham, Pargal, Moreau, Sandberg, Yuan,
  Kushari \& Sanjose]{shubham2023data}
{\sc \au{Shubham, S.}, \au{Pargal, S.}, \au{Moreau, S.}, \au{Sandberg, R.~D.},
  \au{Yuan, J.}, \au{Kushari, A.} \& \au{Sanjose, M.}} \yr{2023} Data-driven
  empirical wall pressure spectrum models for fan noise prediction.  \bt{In
  {\em AIAA AVIATION 2023 Forum\/}},  \pg{p. 3508}.

\bibitem[Simpson {\em et~al.\/}(1987)Simpson, Ghodbane \&
  McGrath]{simpson1987surface}
{\sc \au{Simpson, R.~L.}, \au{Ghodbane, M.} \& \au{McGrath, B.~E.}} \yr{1987}
  \at{Surface pressure fluctuations in a separating turbulent boundary layer}.
  \jt{Journal of Fluid Mechanics}  \bvol{177},  \pg{167--186}.

\bibitem[Swanepoel {\em et~al.\/}(2023)Swanepoel, Biedermann \& van~der
  Spuy]{swanepoel2023experimental}
{\sc \au{Swanepoel, P.~C.}, \au{Biedermann, T.~M.} \& \au{van~der Spuy, S.~J.}}
  \yr{2023}  \at{Experimental noise reduction (aeroacoustical enhancement) of a
  large diameter axial flow cooling fan through a reduction in blade tip
  clearance}.  \jt{International Journal of Aeroacoustics}  \pg{p.
  1475472X231183156}.

\bibitem[Thomson \& Rocha(2022)]{thomson2022semi}
{\sc \au{Thomson, N.} \& \au{Rocha, J.}} \yr{2022}  \at{Semi-empirical wall
  pressure spectral modeling for zero and favorable pressure gradient flows}.
  \jt{The Journal of the Acoustical Society of America}  \bvol{152}~(1),
  \pg{80--98}.

\bibitem[Van~Blitterswyk \& Rocha(2017)]{van2017experimental}
{\sc \au{Van~Blitterswyk, J.} \& \au{Rocha, J.}} \yr{2017}  \at{{An
  experimental study of the wall-pressure fluctuations beneath low Reynolds
  number turbulent boundary layers}}.  \jt{The Journal of the Acoustical
  Society of America}  \bvol{141}~(2),  \pg{1257--1268}.

\bibitem[Venkatraman {\em et~al.\/}(2023)Venkatraman, Moreau, Christophe \&
  Schram]{venkatraman2023numerical}
{\sc \au{Venkatraman, K.}, \au{Moreau, S.}, \au{Christophe, J.} \& \au{Schram,
  C.~F.}} \yr{2023} {Numerical investigation of aerodynamics and aeroacoustics
  of helical Darrieus wind turbines}.  \bt{In {\em AIAA AVIATION 2023
  Forum\/}},  \pg{p. 3641}.

\bibitem[Willmarth(1975)]{willmarth1975pressure}
{\sc \au{Willmarth, W.~W.}} \yr{1975}  \at{Pressure fluctuations beneath
  turbulent boundary layers}.  \jt{Annual Review of Fluid Mechanics}
  \bvol{7}~(1),  \pg{13--36}.

\bibitem[Wu {\em et~al.\/}(2019)Wu, Moreau \& Sandberg]{wu2019effects}
{\sc \au{Wu, H.}, \au{Moreau, S.} \& \au{Sandberg, R.~D.}} \yr{2019}
  \at{{Effects of pressure gradient on the evolution of velocity-gradient
  tensor invariant dynamics on a controlled-diffusion aerofoil at
  $Re_c=150000$}}.  \jt{Journal of Fluid Mechanics}  \bvol{868},
  \pg{584--610}.

\bibitem[Wu \& Piomelli(2018)]{wu2018effects}
{\sc \au{Wu, W.} \& \au{Piomelli, U.}} \yr{2018}  \at{Effects of surface
  roughness on a separating turbulent boundary layer}.  \jt{Journal of Fluid
  Mechanics}  \bvol{841},  \pg{552--580}.

\end{thebibliography}

\end{document}